\documentstyle[prd,aps,preprint,psfig,floats]{revtex}
\begin{document}
\newcommand{\eV}{~\mbox{eV}}
\newcommand{\keV}{~\mbox{keV}}
\newcommand{\MeV}{~\mbox{MeV}}
\newcommand{\GeV}{~\mbox{GeV}}
\newcommand{\TeV}{~\mbox{TeV}}
\newcommand{\for}{~~\mbox{for}~}
\newcommand{\fn}{~~\mbox{for}~n=1}
\preprint{\begin{tabular}{l}
\hbox to\hsize{\hfill UT-855}\\
\hbox to\hsize{\hfill RESCEU-28/99}\\
\hbox to\hsize{\hfill}\\
\end{tabular}}
\title{Leptogenesis in Inflationary Universe}
\author{T. Asaka,$^1$ K. Hamaguchi,$^1$
M. Kawasaki$^2$ and T. Yanagida$^{1,2}$}
\address{$^1$Department of Physics,  University of
  Tokyo, Tokyo 113-0033, Japan}
\address{$^2$Research Center for the Early Universe,
  University of Tokyo, Tokyo 113-0033, Japan}
\date{July 30, 1999}

\maketitle
\begin{abstract}
    We investigate the leptogenesis via decays of heavy Majorana
    neutrinos which are produced non-thermally in inflaton decays. We
    make a comprehensive study on the leptogenesis assuming various
    supersymmetric (SUSY) models for hybrid, new and topological
    inflations.  For an estimation of the lepton asymmetry we adopt
    the Froggatt-Nielsen mechanism for mass matrices of quarks and
    leptons. We find that all of these models are successful to
    produce the lepton asymmetry enough to explain the baryon number
    in the present universe. Here we impose low reheating temperatures
    such as $T_R \lesssim 10^8$ GeV in order to suppress the abundance
    of gravitinos not to conflict with the big-bang nucleosynthesis.
    Furthermore, we find that the leptogenesis works very well even
    with $T_R \simeq 10^{6}$ GeV in the SUSY hybrid or new inflation
    model. It is known that such a reheating temperature is low enough
    to suppress the abundance of gravitinos of mass $m_{3/2} \simeq
    100$ GeV--1 TeV.  Thus, the leptogenesis is fully consistent with
    the big-bang nucleosynthesis in a wide region of the gravitino
    mass.
\end{abstract}
\clearpage
%
%
\section{Introduction}
\label{sec:intro}
It was first pointed out by 'tHooft in 1976 that instanton-like
nonperturbative effects violate baryon-number conservation in the
standard electroweak gauge theory \cite{tHooft}.  These effects are
exponentially suppressed by a large factor at zero temperature and
hence the stability of proton is practically guaranteed.  Kuzmin,
Rubakov and Shaposhnikov suggested \cite{Kuzmin-Rubakov-Shaposhnikov},
however, that the baryon-number violation is not suppressed and can be
even efficient at high temperatures above the electroweak phase
transition.  These baryon-number violating (``sphaleron'') processes
conserve a linear combination of baryon ($B$) and lepton ($L$)
numbers, that is, $B-L$.  Therefore, if the baryon asymmetry was
generated in the early universe with $B-L$ conserving processes as in
the standard $SU(5)$ grand unified theory, all baryon asymmetry would
be washed out by the ``sphaleron'' effects
\cite{Fukugita-Yanagida-Sph}.%
\footnote{ This is not the case if the baryon asymmetry is produced
after or during the electroweak phase transition (see Refs.
\cite{Affleck-Dine} and \cite{EW-Baryogenesis}).  }
The existence of baryon (matter-antimatter) asymmetry in the present
universe, thus, indicates that the $B-L$ violating processes 
played an important role on generating the baryon asymmetry 
in the early universe.

It is quite natural to consider that such $B-L$ violating interactions
induce effective operators of $B-L$ violation at low energies.  The
lowest dimensional possible operators in the electroweak gauge theory is
\begin{eqnarray}
    \label{eq:LV-O}
    {\cal O} = \frac{ f_{ij} }{ M } 
    l_i l_j H H~,
\end{eqnarray}
where $l_i$ ($i$=1, 2, 3) and $H$ denote lepton doublets and the Higgs
scalar field, respectively, and $M$ is the scale of $B-L$ violation.
These operators induce small neutrino masses in the vacuum $\langle H
\rangle \simeq$ 246/$\sqrt{2}$ GeV \cite{See-Saw}.  There is now a
convincing experimental evidence \cite{Super-K} that one of neutrinos
has indeed a small mass $m_\nu$ of order 0.1 eV, which would encourage
us to continue the above consideration to some extent.

The lepton (or $B-L$)-number violating operators (\ref{eq:LV-O}) arise
from the exchange of heavy Majorana neutrinos $N_i$ ($i=$ 1, 2, 3)
\cite{See-Saw}.  Decays of these heavy neutrinos $N_i$ produce
very naturally lepton (or $B-L$) asymmetry if $C$ and $CP$ are not
conserved \cite{Fukugita-Yanagida}.  The produced lepton asymmetry is
converted to baryon asymmetry through the ``sphaleron'' effects in the
early universe \cite{Fukugita-Yanagida}.  Therefore, the leptogenesis
seems the most natural mechanism to account for the baryon asymmetry
in the present universe.

There have been, so far, proposed various scenarios for the
leptogenesis depending on production mechanisms of the heavy Majorana
neutrinos $N_i$
\cite{Fukugita-Yanagida,LG-thermal,LG-infdecay,LG-osc}.  Among them
the most conventional is the thermal production of $N_i$ in the early
universe.  The detail analysis \cite{LG-thermal} shows that sufficient
lepton asymmetry to explain the present baryon asymmetry can be
obtained if the reheating temperature $T_R$ of inflation is ${\cal
O}(10^{10})$ GeV. However, when we consider the leptogenesis in the
framework of supergravity, we have a cosmological gravitino problem:
too many gravitinos are produced with $T_R = {\cal
O}(10^{10})$ GeV to maintain the success of 
the big-bang nucleosynthesis (BBN)
\cite{Gravitino-Prob}.  One way to solve the problem is to assume that
the gravitino is the lightest supersymmetry (SUSY) particle of mass
$m_{3/2} \simeq 10$--$100$ GeV
\cite{Bolz-Buchmuller-Plumacher}.%
\footnote{
Another solution is to consider 
$m_{3/2} \lesssim 1$ keV \cite{Pagels-Primack}.
}
Although this solution is a consistent and even interesting
possibility, it is very important to find other leptogenesis scenarios
consistent with the BBN in a wide region of the gravitino mass
$m_{3/2} \simeq 100$ GeV--1 TeV.

In a recent article \cite{Asaka-Hamaguchi-Kawasaki-Yanagida} we
discussed leptogenesis via decays of the heavy Majorana neutrinos
produced non-thermally in inflaton decays and found that the desired
amount of lepton asymmetry is obtained for $T_R \simeq 10^8$ GeV. In
this case the gravitino of mass $m_{3/2} \simeq$ 500GeV--1 TeV becomes
consistent with the BBN \cite{Gravitino-Prob}.  The purpose of this
paper is to perform a detailed analysis on the leptogenesis, adopting
a larger class of inflation models.  We find that the required lepton
asymmetry $n_L/s \simeq - 10^{-10}$ (the ratio of the lepton-number
density $n_L$ to the entropy density of the universe $s$) can be
obtained even for $T_R \simeq 10^{6}$ GeV in some of the inflation
models, and hence we conclude that there is no cosmological gravitino
problem in the interesting wide region of the gravitino mass $m_{3/2}
\simeq 100$ GeV--1 TeV.  We assume SUSY and hence supergravity
throughout this paper.

In Sec.~\ref{sec:LA-InfDecay} we review briefly the lepton-number
production in the decays of the heavy Majorana neutrinos.  The
magnitude of the lepton-asymmetry parameter $\epsilon_1$ depends on
various Yukawa coupling constants $(h_\nu)_{ij}$ and the mass ratios
$M_i/M_j$ ($M_i$ denote masses of the $N_i$).  For an estimation of
$\epsilon_1$ we adopt the Froggatt-Nielsen model
\cite{Froggatt-Nielsen} for mass matrices of quarks and leptons
including $N_i$, since it is very ``successful'' in explaining the
observed mass hierarchies and mixing angles.  The leptogenesis in a
hybrid inflationary universe is discussed in
Sec.~\ref{sec:LG-HybridInf}, where we consider two different types of
SUSY hybrid inflation models.  In Sec.~\ref{sec:LG-NewInf} we discuss
the leptogenesis in a SUSY new inflation model.  We also consider the
case of a SUSY topological inflation in Sec.~\ref{sec:LG-TopInf}. 
The last section
is devoted to discussion and conclusions.
%
\section{Lepton Asymmetry in Decays of Heavy Majorana Neutrinos}
\label{sec:LA-InfDecay}
%
Before discussing lepton-number generation in decays of the heavy
Majorana neutrinos, we briefly review the Froggatt-Nielsen (FN) model
\cite{Froggatt-Nielsen} for mass matrices of quarks and leptons.  We
adopt the FN model throughout this paper, since it seems the most
attractive mechanism for explaining the observed hierarchies in the
quark and charged lepton mass matrices.  The model is based on a
broken $U(1)_F$ family symmetry.  A gauge-singlet field $\Phi$
carrying the FN charge $Q_\Phi = -1$ is assumed to have a
vacuum-expectation value (vev) $\langle \Phi \rangle \neq 0$ and then
the Yukawa couplings of Higgs supermultiplets arise from
nonrenormalizable interactions of $\Phi$ as
\begin{eqnarray}
    \label{eq:W-FN}
    W = 
    g_{ij}  \Phi^{Q_i + Q_j }
    \Psi_i \Psi_j H_{u(d)}~,
\end{eqnarray}
where $Q_i$ are the FN charges of various supermultiplets
$\Psi_i$, $g_{ij}$ ${\cal O}(1)$ coupling constants.  
$H_u$ and $H_d$ are Higgs
supermultiplets which couple to up-type and down-type quarks,
respectively. Here and
hereafter, we take the gravitational scale $M_G \simeq 2.4 \times
10^{18}$ GeV equal to unity.  The mass hierarchies for quarks and
charged leptons are well explained in terms of their FN charges listed
in Table \ref{tab:FNcharges} with $\epsilon \equiv \langle \Phi
\rangle \simeq 1/17$ \cite{Sato-Yanagida}.
The charges for $e_i^c$ are taken to be the same as those of 
up-type quarks assuming that they belong to the same {\bf 10}'s 
in the $SU(5)$ grand unified theory.
The charge $a$ of the $l_3$ may be 0 or 1
(see Ref. \cite{Sato-Yanagida} for details).
\begin{table}[t]
    \begin{center}
    \begin{tabular}{| c | c c c | c c c  | c c c | }
        $\Psi_i$ & $l_3$ & $l_2$ & $l_1$ 
                 & $e_3^c$ & $e_2^c$ & $e_1^c$ 
                 & $N_3$ & $N_2$ & $N_1$ \\
        \hline
        $Q_i$    & $a$   & $a$   & $a+1$
                 & 0     & 1     & 2
                 & b     & c     & d \\
    \end{tabular}
    \caption{The FN charges of various supermultiplets. 
    We assume $a=0$ or 1 and $b \le c < d$ in the text.}
    \label{tab:FNcharges}
    \end{center}
\end{table}

We apply the above mechanism to the heavy neutrino sector.  Possible FN
charges of the heavy Majorana neutrinos are also found in Table
\ref{tab:FNcharges}.  Then, Majorana masses $M_i$ of the heavy
neutrinos $N_i$ are given by
\begin{eqnarray}
    \label{eq:Mass-Ni}
    M_i \simeq \epsilon^{2 Q_i} M_0 ~,
\end{eqnarray}
where $Q_i$ denote the FN charges of $N_i$ and $M_0$ some mass scale
of the $B-L$ breaking (see discussion in the next section).
Furthermore, with the FN charges of the lepton doublets $l_i$
in Table \ref{tab:FNcharges}
the see-saw mechanism \cite{See-Saw} induces 
a mass matrix for neutrinos
\cite{Sato-Yanagida,Buchmuller-Yanagida} as
\begin{eqnarray}
    \label{mnu}
    (m_{\nu})_{ij}
    \simeq
    \epsilon^{2 a}
    \left( 
        \begin{array}{ c c c}
            \epsilon^2 & \epsilon & \epsilon \\
            \epsilon   & 1        & 1 \\
            \epsilon   & 1        & 1 
        \end{array}
    \right)
    \frac{ \langle H_u \rangle^2 }{ M_0 } ~.
\end{eqnarray}
As shown in Ref. \cite{Sato-Yanagida}, this mass matrix leads to a
large $\nu_\mu$--$\nu_\tau$ mixing angle which is consistent with the
atmospheric neutrino oscillation observed in the Superkamiokande
experiments \cite{Super-K}.  From the observed value of the mass
squared difference for neutrinos $\nu_i$ ($i$ = 1, 2, 3), $m_{\nu_3}^2
- m_{\nu_2}^2 = (0.5$--$6)\times 10^{-3}$ eV$^2$ \cite{Super-K}, we
find the mass of the heaviest neutrino $\nu_3$ as $m_{\nu_3} \simeq$
(2--8) $\times 10^{-2}$ eV, provided a mass hierarchy, $m_{\nu_3} \gg
m_{\nu_2}$.
Then, we may derive%
\footnote{ We assume, in this paper, $\tan \beta \equiv \langle H_u
\rangle / \langle H_d \rangle \simeq 1$ and hence $\langle
H_u \rangle \simeq 123$ GeV.    
Even if one takes a larger $\tan \beta \simeq 50$, the
discussion in the text does not change too much
(i.e., the obtained lepton asymmetry is reduced only by factor 2).  }
\begin{eqnarray}
    \label{eq:M0}
    M_0 &\simeq& \epsilon^{2a} \frac{ \langle H_u \rangle^2 }
    { m_{\nu_3} }
    \simeq
    \epsilon^{2a} (0.3\mbox{--}1)\times10^{15} ~\mbox{GeV} ~
    \simeq
    \left\{
        \begin{array}{ll}
           (0.3\mbox{--}1) \times 10^{15} ~\mbox{GeV} ~~
           &\mbox{for}~~ a = 0\\
           (1\mbox{--}3) \times 10^{12} ~\mbox{GeV} ~~
           &\mbox{for}~~ a = 1
        \end{array}
    \right.~.
\end{eqnarray}

We are now at the point to see generation of the lepton asymmetry
in decays of the heavy Majorana neutrinos.
Because the decays of the heavy Majorana neutrinos $N_i$ into
lepton $l_j$ and Higgs $H_u$ doublets have the following 
two distinct decay channels;
\begin{eqnarray}
    && N_i \rightarrow H_u + l_j \nonumber \\
    && N_i \rightarrow \overline{H_u} + \overline{l_j}~,
\end{eqnarray}
the interference between decay amplitudes of tree and one-loop
diagrams results in the lepton-number asymmetry
if $CP$ is not conserved \cite{Fukugita-Yanagida}.  
Here and hereafter, 
$N_i$, $l_j (\overline{l_j})$ and $H_u (\overline{H_u})$
denote fermionic or scalar (bosonic) components of corresponding
supermultiplets unless we explicitly distinguish them. 
We consider only the $N_1$ decay, provided
that the mass $M_1$ is much smaller than the others ($M_1 \ll M_2,
M_3$, i.e., $d > b,c$).  From Eqs. (\ref{eq:Mass-Ni}) and
(\ref{eq:M0}) the $M_1$ is evaluated with the FN charge $d$ as
\begin{eqnarray}
    \label{eq:Mass-N1}
    M_1 &\simeq& \epsilon^{2d} M_0
    \simeq
    \left\{
        \begin{array}{ll}
           (1\mbox{--}3) \times 10^{12} ~\mbox{GeV} ~~
           &\mbox{for}~~ a + d = 1\\
           (0.3\mbox{--}1) \times 10^{10} ~\mbox{GeV} ~~
           &\mbox{for}~~ a + d = 2
        \end{array}
    \right.~.
\end{eqnarray}
In the following analysis, we only consider the case
$a+d=1$ or $a+d=2$.
The total decay width of the $N_1$,  
$\Gamma_{N_1}$, is given by
\begin{eqnarray}
    \label{eq:Gam-N1}
    \Gamma_{N1} 
    &\simeq &
    \frac{ 1 }{ 4 \pi }
    \epsilon^{ 2(a+d) } M_1
    \simeq
    \left\{
        \begin{array}{ll}
           (0.3\mbox{--}1) \times 10^{9} ~\mbox{GeV} ~~
           &\mbox{for}~~ a + d = 1\\
           (0.3\mbox{--}1) \times 10^{4} ~\mbox{GeV} ~~
           &\mbox{for}~~ a + d = 2
        \end{array}
    \right.~.
\end{eqnarray}
The lepton asymmetry produced in the $N_1$ decay is represented by
a parameter $\epsilon_1$, which is calculated as
\cite{Covi,Buchmuller-Plumacher}
\begin{eqnarray}
    \epsilon_1 
    &\equiv&
    \frac{ \Gamma (N_1 \rightarrow H_u + l )
         - \Gamma (N_1 \rightarrow \overline{H_u} + \overline{l} ) }
         { \Gamma (N_1 \rightarrow H_u + l )
         + \Gamma (N_1 \rightarrow \overline{H_u} + \overline{l} ) }
    \nonumber \\
    &=&
    - 
    \frac{ 3 }{ 16 \pi \left( h_\nu h_\nu^{\dagger} \right)_{11} }
    \left[ 
        \mbox{Im} \left( h_\nu h_\nu^{\dagger} \right)_{13}^2 
        \frac{ M_1 }{ M_3 }
        +
        \mbox{Im} \left( h_\nu h_\nu^{\dagger} \right)_{12}^2 
        \frac{ M_1 }{ M_2 }
    \right]
    .
    \label{Ep1}
\end{eqnarray}
Here we have taken a basis where the mass matrix for $N_i$ is diagonal
and the Yukawa couplings $(h_\nu)_{ij}$ are defined in the
superpotential as $W = (h_\nu)_{ij} N_i l_j H_u$.  We have included
both of one-loop vertex and self-energy corrections
\cite{Buchmuller-Plumacher}.%
\footnote{
$N_1$ in Eq. (\ref{Ep1}) denotes fermionic or scalar component
of the supermultiplet $N_1$. Thus,
the lepton-asymmetry parameter $\epsilon_1$ in the decay of scalar
component $\widetilde{N_1}$ is the same as that in the decay
of fermionic component $N_1$.
}

The FN model [see Eqs. (\ref{eq:W-FN}) and (\ref{eq:Mass-Ni})]
allows us to rewrite the
lepton-asymmetry parameter $\epsilon_1$ in Eq. (\ref{Ep1}) 
by using an effective $CP$-violating phase $\delta_{\rm eff}$ as
\cite{Asaka-Hamaguchi-Kawasaki-Yanagida}
\begin{eqnarray}
    \epsilon_1 &\simeq&
    \frac{ 3 \delta_{\rm eff} }
         { 16 \pi \left( h_\nu h_\nu^{\dagger} \right)_{11} }
    \left| \left( h_\nu h_\nu^\dagger \right)^2_{13} \right|
    \frac{ M_1 }{ M_3 }
    \simeq
    \frac{ 3 \delta_{\rm eff} }{ 16 \pi }
    \epsilon^{2 (a+d)}~,
    \\
    &\simeq&
    \frac{ 3 \delta_{\rm eff} }{ 16 \pi }
    \frac{ m_{\nu_3} M_1 }{ \langle H_u \rangle^2 }~,
    \label{eq:Ep1_LQ}
\end{eqnarray}
where we have also used Eq. (\ref{mnu}) to derive the last equation
(\ref{eq:Ep1_LQ}).  Notice that the asymmetry parameter
$\epsilon_1$ depends only  on the FN charge ($a+d$).  Taking the
maximum $CP$ violating phase $|\delta_{\rm eff}| \simeq 1$, we obtain
\cite{Buchmuller-Yanagida}
\begin{eqnarray}
    \label{eq:VEp1}
    \epsilon_1
    &\simeq&
    \left\{
        \begin{array}{ll}
          - 2 \times 10^{-4}
           &\mbox{for}~~ a + d = 1\\
          - 7 \times 10^{-7}
           &\mbox{for}~~ a + d = 2
        \end{array}
    \right.~,   
    \\
    &\simeq&
    - (1\mbox{--}3) \times 10^{-6}
    \left( \frac{ M_1 }{ 10^{10}~\mbox{GeV} } \right).
\end{eqnarray}

Let us now assume that the heavy neutrinos $N_1$ are produced
in inflaton $\varphi$ decay, which leads to a constraint 
on the inflaton mass $m_\varphi$ as
\begin{eqnarray}
    \label{eq:const-mphi}
    m_\varphi > 2 M_1,
\end{eqnarray}
and calculate a net lepton-number asymmetry created via the decays of
$N_1$. First of all, we restrict ourselves only to inflation models
with reheating temperatures $T_R$ being lower than $10^8$ GeV to
avoid overproduction of the gravitinos.  In this case the produced
heavy Majorana neutrinos $N_1$ are always out of thermal equilibrium
since $M_1 \gtrsim 10 T_R$ [see Eq. (\ref{eq:Mass-N1}) and discussions in
subsequent sections] and the $N_1$ behave like frozen-out,
relativistic particles with the energy of $E_{N_1} = m_\varphi/2$.
Furthermore, the $N_1$ decay immediately after produced in the
inflaton $\varphi$ decay as we will see in the subsequent sections.
With these conditions we may easily estimate lepton-to-entropy ratio
as \cite{LG-infdecay}
\begin{eqnarray}
    \frac{ n_L }{s} 
    &\simeq&
    \frac{ 3 }{ 2 } 
    \epsilon_1 B_r
    \frac{ T_R }{ m_\varphi } 
    \nonumber \\
    &\simeq& 
    - (1\mbox{--}3) \times 10^{-6} B_r
    \left( \frac{ T_R }{ 10^{10} ~\mbox{GeV} } \right)
    \left( \frac{ M_1 }{ m_\varphi } \right),
    \label{LA}
\end{eqnarray}
where $B_r$ is the branching ratio of the inflaton decay 
$\varphi \rightarrow N_1 N_1$ channel.
Notice that the reheating temperature $T_R$ is bounded from below,
$T_R \gtrsim 10^{5}$ GeV, otherwise the produced lepton asymmetry
is too small to explain the baryon number in the present universe.

The lepton asymmetry in Eq. (\ref{LA}) is converted to the baryon
asymmetry through the ``sphaleron'' effects which is given by 
\begin{eqnarray}
    \frac{ n_B }{ s } \simeq a \frac{ n_L }{ s },
\end{eqnarray}
with $a \simeq - 8/23$ \cite{Khlebnikov-Shaposhnikov} in the minimal 
SUSY standard model. 
To explain the observed baryon asymmetry 
\begin{eqnarray}
    \frac{ n_B }{ s } 
    \simeq ( 0.1 \mbox{--} 1 ) \times 10^{-10},
\end{eqnarray}
we should have the lepton asymmetry
\begin{eqnarray}
    \label{eq:LA-Req}
    \frac{n_L}{ s} \simeq - (0.3\mbox{--}3) \times 10^{-10}.
\end{eqnarray}
In the subsequent sections \ref{sec:LG-HybridInf}, \ref{sec:LG-NewInf}
and \ref{sec:LG-TopInf} we will examine whether the required lepton
asymmetry Eq.~(\ref{eq:LA-Req}) is obtained with low enough reheating
temperature of  $T_R \lesssim 10^{8}$ GeV to avoid the
cosmological gravitino problem. For practical calculations we will use
three types of SUSY models for hybrid, new and topological inflations.
\section{Leptogenesis in Hybrid Inflation}
\label{sec:LG-HybridInf}
In this section we perform a detailed analysis on hybrid inflation
models and examine whether they can provide us with sufficient lepton
asymmetry to account for the baryon asymmetry in the present universe,
avoiding overproduction of the gravitinos to maintain the success of
the BBN. Before discussing hybrid inflation models, we show first a
particle-physics model for the heavy Majorana neutrinos $N_i$.

A simple extension of the SUSY standard electroweak gauge theory 
is given by considering the gauged $B-L$ symmetry,
in which right-handed neutrinos $N_i$  are 
necessary to cancel $B-L$ gauge anomaly.
We introduce standard-model gauge-singlet supermultiplets
$\Psi (x, \theta)$ and $\overline{\Psi} (x, \theta)$ carrying
$B-L$ charges +2 and $-2$, respectively,
and suppose that the $B-L$ symmetry is spontaneously broken by
the condensations $\langle \Psi \rangle$ = $\langle \overline{\Psi}
\rangle$ at high energies.%
\footnote{
We always take $\Psi = \overline{\Psi}$ to satisfy the $D$-term
flatness condition of the $U(1)_{B-L}$.
}
The heavy neutrinos $N_i$ acquire Majorana masses through the following
superpotential;%
\footnote{
Lazarides has discussed \cite{Lazarides} the leptogenesis 
in the hybrid inflation
assuming the nonrenormalizable superpotential 
$W = (g'_i/2) N_i N_i \Psi \Psi$,
where the $B-L$ charge of $\Psi$ is taken as $+1$.
}
\begin{eqnarray}
    \label{eq:W-BL}
    W= \frac{1}{2} g_i N_i N_i \Psi ~.
\end{eqnarray}
Here we have taken a basis where the above Yukawa coupling 
matrix $g$ is diagonal.
We assume that $\Psi$ and $\overline{\Psi}$ have zero FN charges and 
then the $M_0$ in Eq. (\ref{eq:M0}) is 
\begin{eqnarray}
    \label{eq:M0-BL}
    M_0 \simeq \langle \Psi \rangle = \langle \overline{ \Psi } \rangle
    ~.
\end{eqnarray}
Namely, the $B-L$ gauge symmetry is broken down
at the scale about $10^{15}$ GeV or $10^{12}$ GeV,
generating large Majorana masses of the right-handed neutrinos
$N_i$ [see Eq. (\ref{eq:M0})].%
\footnote{
This model is easily embedded in the $SO(10)$ grand unified theory.
}

A superpotential causing the $B-L$ breaking is given by
\begin{eqnarray}
    \label{eq:W-HInf}
    W = - \mu^2 \phi + \lambda \phi \Psi \overline{\Psi} ~,
\end{eqnarray}
where $\phi (x, \theta)$ is a gauge-singlet supermultiplet,
$\lambda$ a coupling constant and $\mu$ a dimensional mass
parameter.
Notice that this superpotential possesses a $U(1)$ $R$-symmetry where 
the $\phi$ and $\Psi \overline{\Psi}$ have $U(1)_R$ charges
2 and 0, respectively.
The potential for scalar components of the supermultiplets 
$\phi(x, \theta)$, $\Psi(x,\theta)$ and $\overline{\Psi}(x,\theta)$
is given by, in supergravity,
\begin{eqnarray}
    \label{V-SUGRA}
    V =
    e^K
   \left\{
  \left(
   \frac{\partial^2 K}{\partial z_I\partial z_J^*}
   \right)^{-1}
   D_{z_I}W D_{z_J^*}W^*
   -
   3|W|^2
   \right\}~+~\mbox{$D$-terms}~,
\end{eqnarray}
where
\begin{eqnarray}
 D_{z_I}W =
  \frac{\partial W}{\partial z_I}
  +
  \frac{\partial K}{\partial z_I} W ~,
\end{eqnarray}
and $z_I$ denote scalar components $\phi$, $\Psi$ and
$\overline{\Psi}$ of the corresponding supermultiplets.
Here we have assumed the $R$-invariant K\"ahler potential
for $\phi$, $\Psi$ and $\overline{\Psi}$ as
\begin{eqnarray}
    \label{eq:K-HInf}
    K = | \phi |^2 +
    |\Psi|^2 + | \overline{\Psi} |^2
    + \frac{ \kappa_1}{4 } | \phi |^4 
    +
    \cdots ~,
\end{eqnarray}
where the ellipsis denotes higher-order terms which we neglect in the
present analysis.
Then, we have a SUSY-invariant vacuum%
\footnote{
We take a basis where $\mu^2$ and $\lambda$ are real and positive by
using the phase rotations of $\phi$ and $\Psi \overline{\Psi}$.
Using the $B-L$ rotation we choose $\langle \Psi \rangle$ to be real
and positive.
}
\begin{eqnarray}
    \label{eq:VEV-HInf}
    &&\langle \Psi \rangle = \langle \overline{\Psi} \rangle
    = \sqrt{ \frac{ \mu^2 }{ \lambda } }  
    ~,
    \nonumber \\
    && \langle \phi \rangle = 0 ~.
\end{eqnarray}

It is quite interesting to observe that the superpotential
(\ref{eq:W-HInf}) is nothing but one proposed in Refs.
\cite{Dvali,Copeland} for a SUSY hybrid inflation model. The real part
of $\phi$ is identified with the inflaton field $\varphi / \sqrt{2}$.
Furthermore, the potential is minimized at $\Psi = \overline{\Psi}=0$
when $\varphi$ is larger than $\varphi_c \equiv \sqrt{ 2 \mu^2 /
\lambda}$, and hybrid inflation occurs for $\varphi_c < \varphi
\lesssim 1$ and $k \equiv - \kappa_1 \ge 0$.  Including one-loop
corrections \cite{Dvali} the potential for the inflaton $\varphi$ is
given by, for $\varphi > \varphi_c$,
\begin{eqnarray}
    \label{eq:V-HInf}
    V &\simeq& \mu^4 + \frac{ k }{ 2 } \mu^4 \varphi^2
    + \frac{1}{16} 
      \left( 4 k^2 + 7 k + 2  \right) \mu^4 \varphi^4
    \nonumber \\
    &&+ \frac{ \lambda^4 }{ 128 \pi^2 }
    \left[ 2 \varphi_c^4\ln 
        \left( \frac{ \lambda^2 \varphi^2 }{ 2 \Lambda^2 } \right)
         + \left( \varphi^2 - \varphi_c^2 \right)^2
         \ln \left( 1 - \frac{\varphi_c^2}{\varphi^2} \right)
         + \left( \varphi^2 + \varphi_c^2 \right)^2
         \ln \left( 1 + \frac{\varphi_c^2}{\varphi^2} \right)
    \right]~,
\end{eqnarray}
where $\Lambda$ denotes the renormalization scale.  Here we have
included in the inflaton potential $\mu^4 \varphi^4$ terms induced by
the K\"ahler potential (\ref{eq:K-HInf}) since the initial value of
the inflaton
field is close to 1.
\footnote{
We have neglected the $\mu^4 \varphi^4$ terms coming from the higher order
interactions in the K\"ahler potential (e.g., $K = \kappa' |\phi|^6$).
}
The mass of the inflaton $\varphi$ in the
true vacuum Eq. (\ref{eq:VEV-HInf}) is estimated as
\begin{eqnarray}
    \label{eq:MPHI-HInf}
    m_{\varphi} \simeq \sqrt{2 \lambda} \mu.
\end{eqnarray}

Let us now discuss the inflation dynamics.
The slow-roll conditions for inflation are given by
\cite{Kolb-Turner},
\begin{eqnarray}
    &&\frac{1}{2} \left( \frac{V'}{V} \right)^2 < 1,
    \label{eq:SR-Cond1}
    \\
    &&\left| \frac{ V''}{V} \right| < 1,
    \label{eq:SR-Cond2}
\end{eqnarray}
where the prime denotes the derivative with the inflaton field.  These
conditions are satisfied when $\varphi > \varphi_c$, $\lambda
< 1$ and $1 > k \ge 0$.  Therefore, while the inflaton
$\varphi$ rolls down along the potential (\ref{eq:V-HInf}) from 
$\varphi_I$ ($1\gtrsim \varphi_I > \varphi_c$) to $\varphi_c$, the vacuum
energy $\mu^4$ of the potential dominates the energy of the universe
and hence the hybrid inflation takes place \cite{Hybrid-Inf-Model}.

After the inflation ends, the vacuum energy is transferred 
into the energies of the coherent oscillations of the inflaton $\varphi$
and the scalar field $\Sigma = (\Psi + \overline{\Psi})/\sqrt{2}$.
The radiations of the universe are produced by the decays of the
$\varphi$ and/or $\Sigma$ field 
and the universe is reheated at the temperature $T_R$.
In order to estimate the reheating temperature $T_R$ we have to know
total decay rates of these scalar fields.
Through the interactions in the superpotentials 
(\ref{eq:W-BL}) and (\ref{eq:W-HInf}),
the inflaton $\varphi$ decays into scalar components 
$\widetilde{N_1}$ of the $N_1$ supermultiplet,
if kinematically allowed, with the rate
\begin{eqnarray}
    \Gamma_\varphi \simeq 
    \Gamma ( \varphi \rightarrow \widetilde{N_1}\widetilde{N_1} )
    =
    \frac{ 1 }{ 64 \pi }
    \frac{ M_1^2 m_\varphi }{ \langle \Psi \rangle^2 }
    \left( 1 - \frac{ 4 M_1^2 }{ m_\varphi^2 } \right)^{1/2}.
\end{eqnarray}
The $\Sigma$ field decays into scalar and fermionic components
of the $N_1$ with
\begin{eqnarray}
    \label{eq:GAMW1-HInf}
    \Gamma (\Sigma \rightarrow {\widetilde N}_1 {\widetilde N}_1) 
    &=& 
    \frac{ 1 }{ 16 \pi }
    \frac{ M_1^4 }{ \langle \Psi \rangle^2 m_\Sigma }
    \left( 1 - \frac{ 4 M_1^2 }{ m_\Sigma^2 } \right)^{1/2} ~,\\
    \Gamma (\Sigma \rightarrow N_1 N_1) 
    &=& 
    \frac{ 1 }{ 64 \pi }
    \frac{ M_1^2 m_\Sigma }{ \langle \Psi \rangle^2 }
    \left( 1 - \frac{ 4 M_1^2 }{ m_\Sigma^2 } \right)^{3/2} ~,
    \label{eq:GAMW2-HInf}
\end{eqnarray}
if $m_\Sigma > 2 M_1$. Notice that the inflaton $\varphi$ forms
a massive supermultiplet together with the $\Sigma$ field 
in the vacuum (\ref{eq:VEV-HInf}) and
the mass of the $\Sigma$ field is
equal to the inflaton mass ($m_\Sigma = m_\varphi \simeq \sqrt{2
\lambda} \mu)$.  

Since the $\Sigma$ field has a non-zero
vev, the $\Sigma$ decays also through nonrenormalizable interactions
in the K\"ahler potential,
\footnote{ The decays of the $\Sigma$ field through the $B-L$ 
gauge-multiplet exchanges are negligible.  }
\begin{eqnarray}
    \label{eq:KTR-HInf}
    K = \sum_i c_i | \Sigma |^2  | \psi_i |^2,
\end{eqnarray}
where $\psi_i$ denote supermultiplets of the SUSY standard-model
particles including the heavy Majorana neutrinos, and $c_i$ coupling
constants
of order unity.  Then the decay rate by these interactions is
estimated as
\begin{eqnarray}
    \label{eq:GAMK-HInf}
    \Gamma (\Sigma \rightarrow \psi_i \overline{\psi_i}) 
    \simeq \frac{1}{8 \pi } C
    \langle \Psi \rangle^2 m_\Sigma^3,
\end{eqnarray}
where $C = \sqrt{ \sum c_i^2 }$ is a parameter of order unity.  In the
following analysis we take $C=1$.  Comparing the rate
(\ref{eq:GAMK-HInf}) with Eqs.  (\ref{eq:GAMW1-HInf}) and
(\ref{eq:GAMW2-HInf}), we easily see that the total decay rate of the
$\Sigma$ field is determined by Eq.  (\ref{eq:GAMW2-HInf}) since
$\Gamma_\Sigma \simeq \Gamma (\Sigma \rightarrow N_1 N_1) > \Gamma
(\Sigma \rightarrow {\widetilde N}_1 {\widetilde N}_1)$ for $m_\Sigma
> 2 M_1$.

We assume $m_\varphi = m_\Sigma > 2 M_1$ as discussed in
the previous section [see Eq. (\ref{eq:const-mphi})], 
and hence the decays of
the inflaton $\varphi$ and the $\Sigma$ field take place at the almost
same time because of $\Gamma_\varphi \simeq \Gamma_\Sigma$.
Thus, the reheating temperature $T_R$ is estimated 
by the total decay width of the $\Sigma$ field as
\begin{eqnarray}
    \label{eq:TR-HInf}
    T_R \simeq 0.46 \sqrt{ \Gamma_\Sigma}~.
\end{eqnarray}
The total decay rate of the $\Sigma$ field 
is much smaller than the decay rate of the $N_1$ [see
Eq. (\ref{eq:Gam-N1})] and the $N_1$ decays immediately after
produced in the $\Sigma$ decay. This guarantees the validity
of the formula of the lepton asymmetry given by Eq. (\ref{LA}).
Here it should be noted that
the branching ratio of the $\Sigma$ (inflaton $\varphi$) decay into two
$N_1$ ($\widetilde{N_1}$) in Eq. (\ref{LA}) is $B_r \simeq1$. 

The above hybrid inflation must explain the following two
observations: (i) the $e$-hold number $N_e$ of the present horizon and
(ii) the density fluctuations observed by the cosmic background
explorer (COBE) satellite.  While the inflaton $\varphi$ rolls down
along the potential (\ref{eq:V-HInf}) from $\varphi_{N_e}$
to $\varphi_c$, the scale factor of the universe
increases by $e^{N_e}$.  This $e$-fold number $N_e$ is given by
\begin{eqnarray}
    \label{eq:Ne}
    N_e \simeq \int_{\varphi_c}^{\varphi_{N_e}} d \varphi ~
    \frac{ V(\varphi) }{ V'(\varphi)} ~.
\end{eqnarray}
In order to explain the present horizon scale, the $e$-fold number
should be 
\begin{eqnarray}
    \label{eq:Ne-PH}
    N_e = 67 + \frac{1}{3} \ln \left( H_I T_R \right),
\end{eqnarray}
where $H_I \simeq \mu^2 /\sqrt{3}$ denotes the Hubble parameter
during the inflation.
Furthermore, the amplitude of the primordial density fluctuations
$\delta \rho / \rho$ predicted by the hybrid inflation,
\begin{eqnarray}
    \frac{ \delta \rho }{ \rho }
    \simeq \frac{ 1 }{ 5 \sqrt{3} \pi }
    \frac{ V^{3/2} (\varphi_{N_e} ) }
         {\left| V'(\varphi_{N_e} )\right|},
\end{eqnarray}
should be normalized by the data on anisotropies of the cosmic 
microwave background radiation (CMBR) observed by the 
COBE satellite \cite{COBE},
which gives
\begin{eqnarray}
    \label{eq:COBE-Norm}
    \frac{ V^{3/2} (\varphi_{N_e} ) }
         {\left| V'(\varphi_{N_e} )\right|}
    \simeq 5.3 \times 10^{-4} ~.
\end{eqnarray}

\begin{figure}[ht]
    \centerline{\psfig{figure=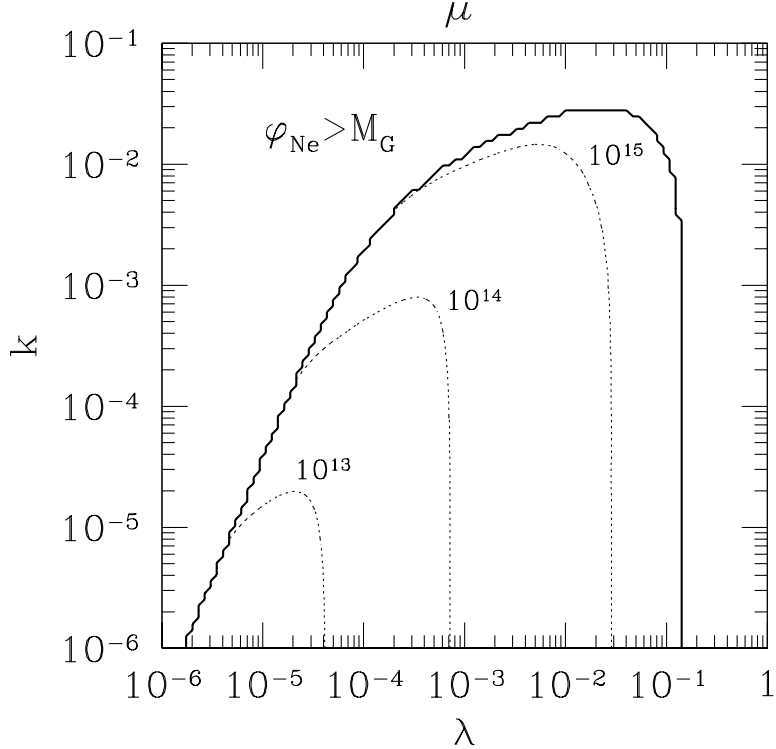,width=12cm}}
    \caption{
    The contour lines of the scale $\mu$ in the hybrid inflation
    model.  The contour lines are all shown by the dotted lines and
    corresponding values of $\mu$ are also represented in unit of GeV.
    The upper bound on $k$ from the requirement
    $\varphi_{N_e} < M_G$ is shown by the thick solid line.
    }
    \label{fig:MU_BL}
\end{figure}

\begin{figure}[ht]
    \centerline{\psfig{figure=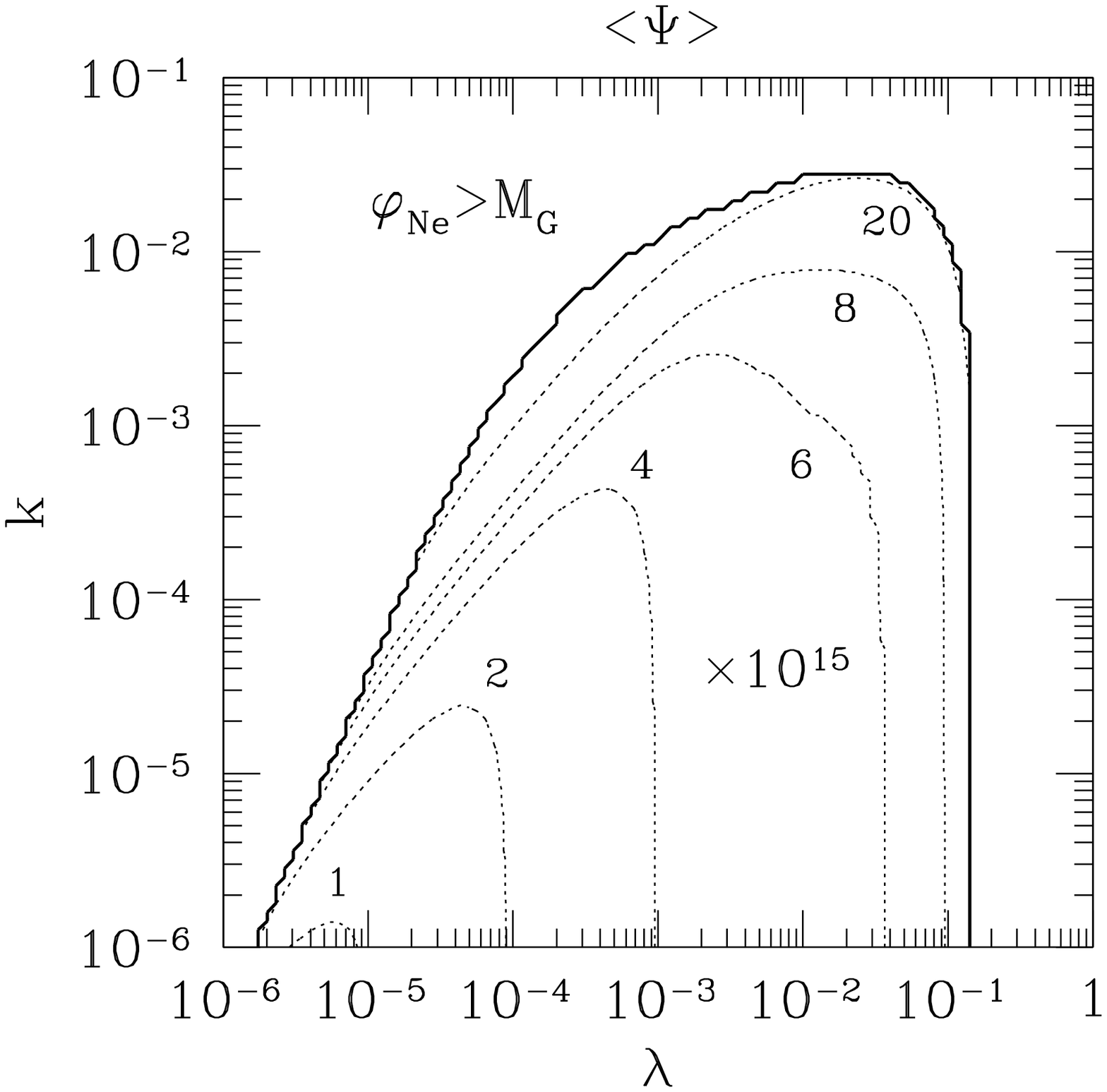,width=12cm}}
    \caption{
    The contour lines of the vev $\langle \Psi \rangle$ in 
    the hybrid inflation model.
    The contour lines are all shown by the dotted lines
    and corresponding values of $\langle \Psi \rangle $ 
    are also represented in unit of  $10^{15}$ GeV.
    The upper bound on $k$ from the requirement $\varphi_{N_e} < M_G$ is 
    shown by the thick solid line.
    }
    \label{fig:VEV_BL}
\end{figure}

\begin{figure}[ht]
    \centerline{\psfig{figure=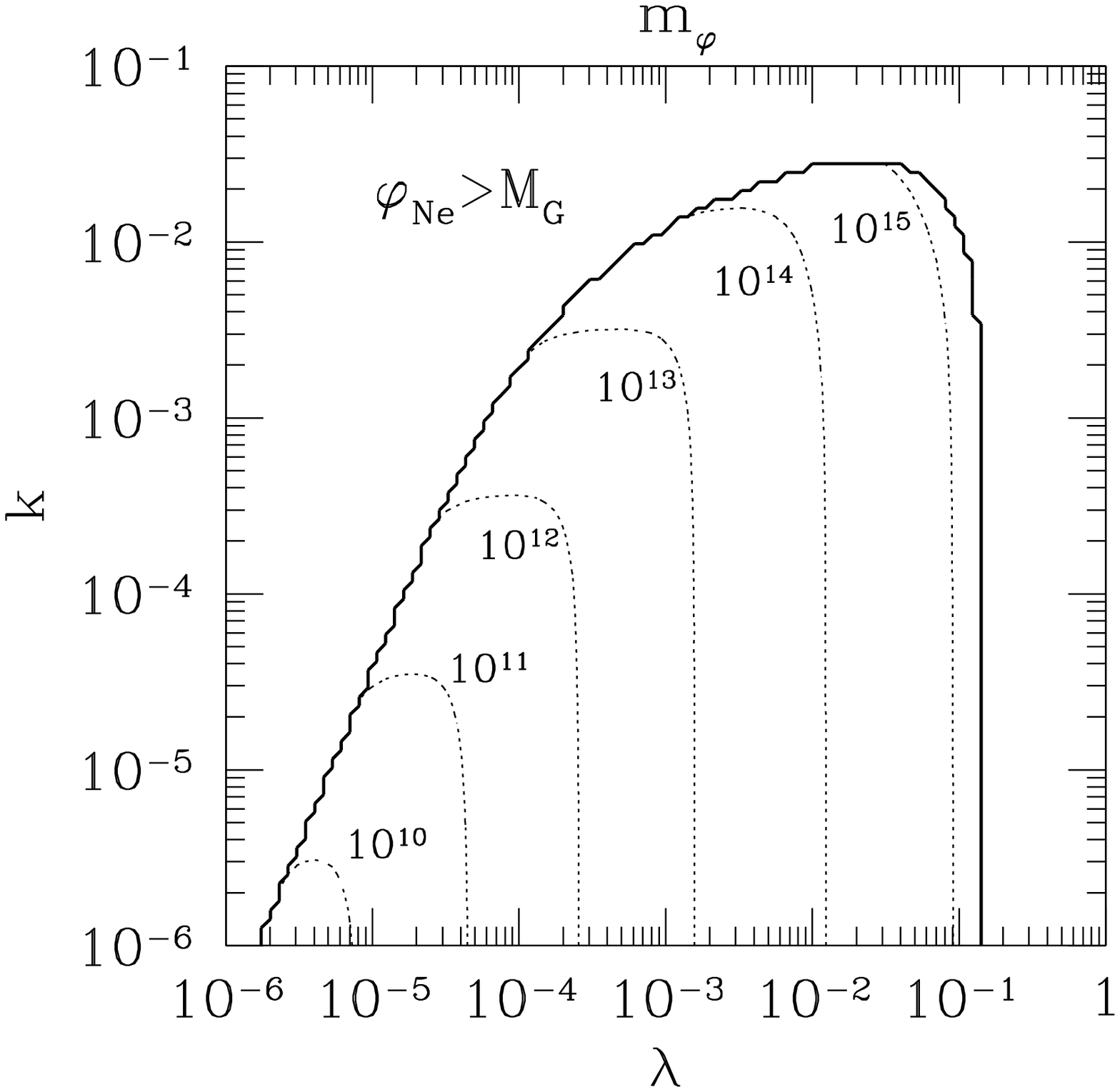,width=12cm}}
    \caption{
    The contour lines of the inflaton mass  $m_\varphi$ in
    the hybrid inflation model.
    The contour lines are all shown by the dotted lines
    and corresponding values of $m_\varphi$ are also represented
    in unit of GeV.
    The upper bound on $k$ from the requirement $\varphi_{N_e} < M_G$ is 
    shown by the thick solid line.
    }
    \label{fig:MPHI_BL}
\end{figure}

From Eqs. (\ref{eq:Ne-PH}) and (\ref{eq:COBE-Norm}) the scale $\mu$ of
the hybrid inflation is determined for given $\lambda$ and $k$.  We
have performed numerical calculations and the obtained $\mu$ is shown
in Fig.  \ref{fig:MU_BL}.  (Note that we find no sizable difference in
the scale $\mu$ for between $a+d=1$ and $a+d=2$.)  We should exclude
the region where $\varphi_{N_e} \gtrsim 1$, because our effective
treatment of the inflaton potential (\ref{eq:V-HInf}) becomes invalid
in that region.
The vev of $\Psi$ and the inflaton mass $m_\varphi$ are found 
in Figs. \ref{fig:VEV_BL} and \ref{fig:MPHI_BL}, respectively.
%
It is interesting that as shown in Fig.~\ref{fig:VEV_BL} the scale of
the $B-L$ breaking is predicted as $\langle \Psi \rangle \simeq$
$(1$--$5)\times10^{15}$ GeV in a wide parameter region, $10^{-6}
\lesssim \lambda \lesssim 10^{-2}$ and $k \lesssim 10^{-3}$, which is
very consistent with $M_0 \simeq 10^{15}$ GeV 
(i.e., $a=0$) derived from the observed
neutrino mass [see Eq. (\ref{eq:M0})].  On the other hand, the lower
value of the $B-L$ breaking scale of $M_0 \simeq 10^{12}$ GeV 
($a=1$) cannot
be obtained in the present hybrid inflation model.

\begin{figure}[ht]
    \centerline{\psfig{figure=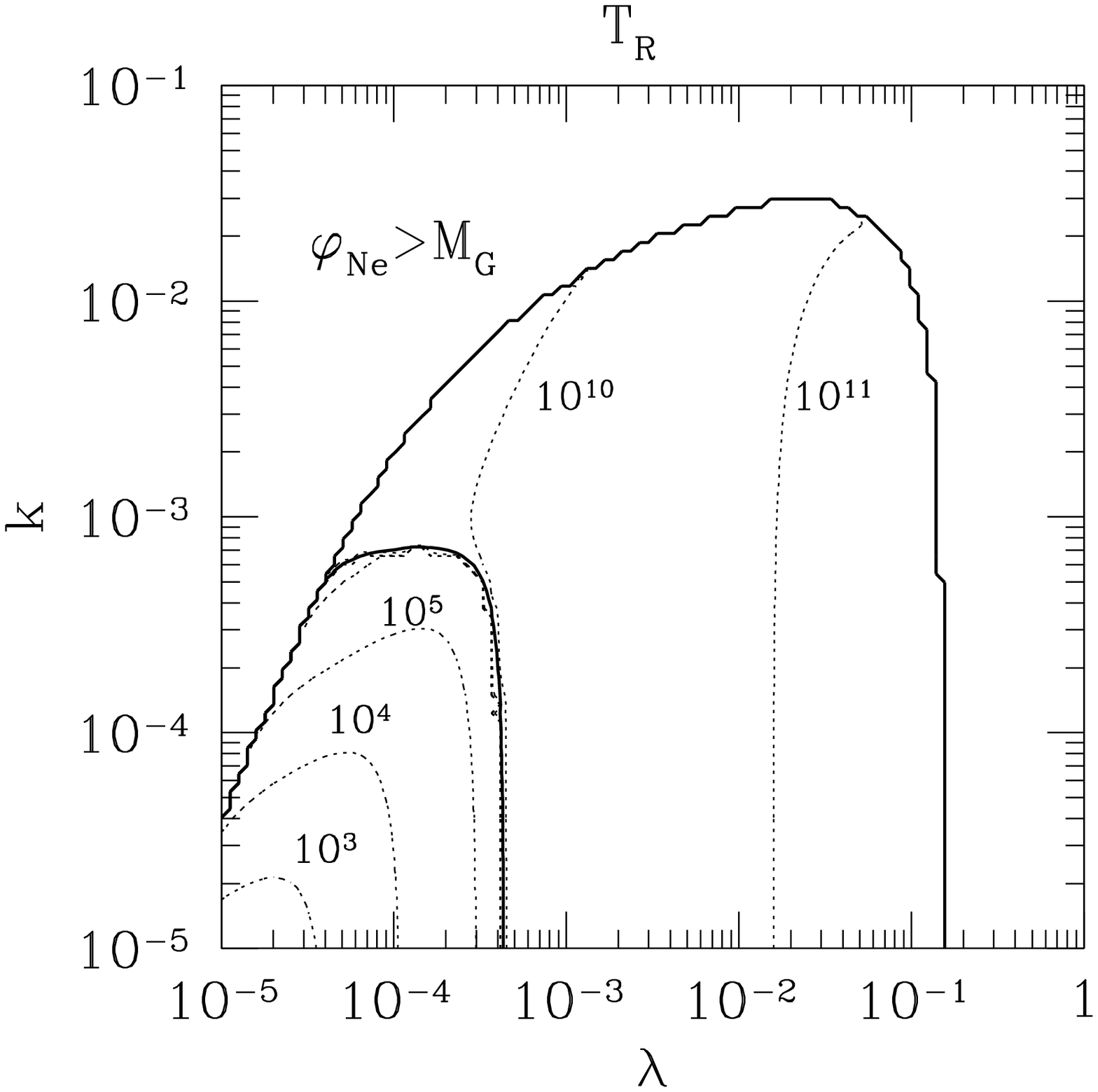,width=12cm}}
    \caption{
    The contour lines of the reheating temperature $T_R$ in
    the hybrid inflation model for the case 
    $M_1 \simeq 10^{12}$ GeV ($a+d=1$).
    The contour lines are all shown by the dotted lines
    and corresponding values of $T_R$ are also represented
    in unit of GeV.
    The upper bound on $k$ from the requirement $\varphi_{N_e} < M_G$ is 
    shown by the thick solid line.
    The $k$ yielding  $m_\varphi = m_\Sigma = 2M_1$ is also shown
    by the thick solid line.
    }
    \label{fig:TR_BL}
\end{figure}

\begin{figure}[t]
    \centerline{\psfig{figure=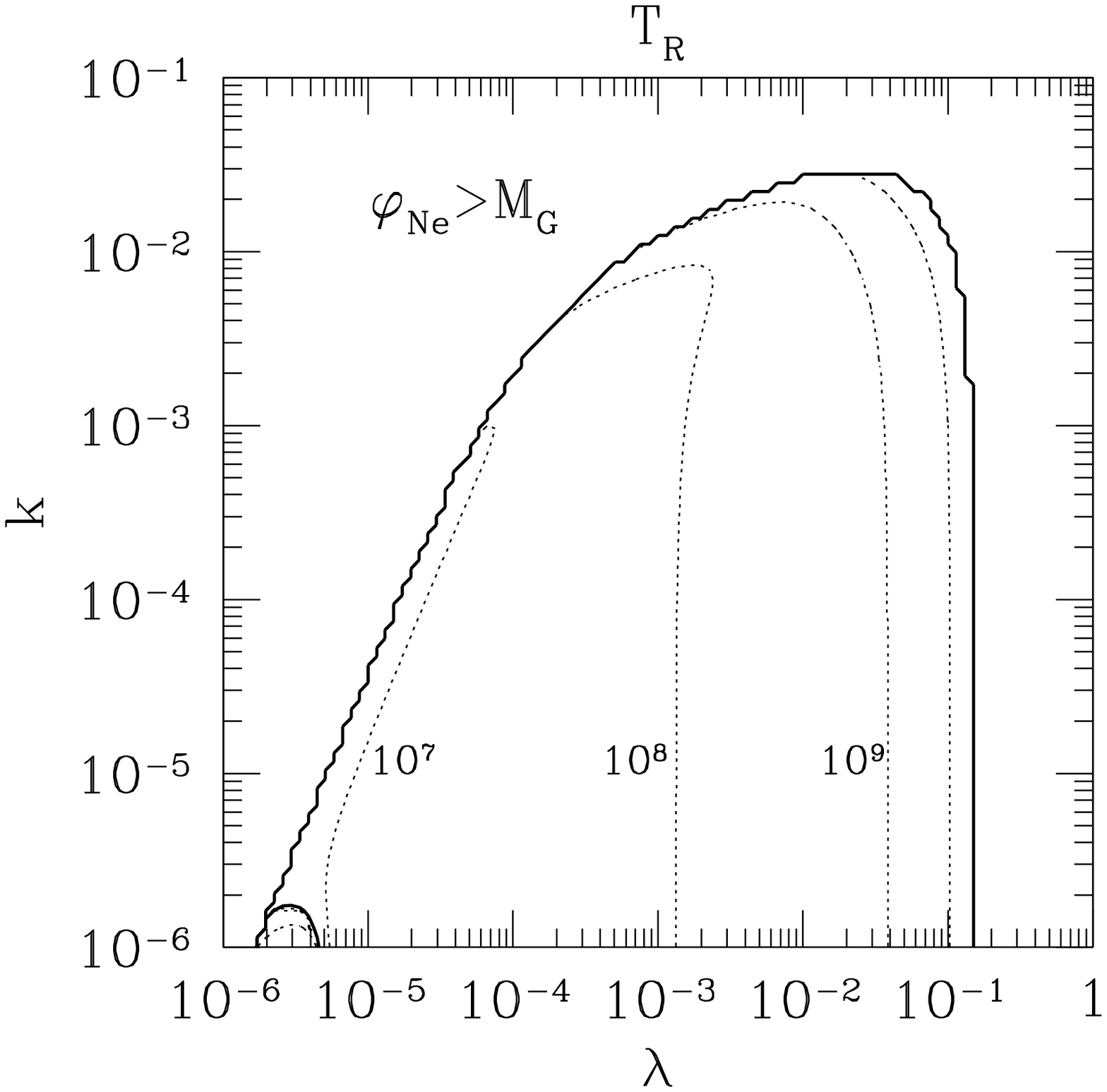,width=12cm}}
    \caption{
    The contour lines of the reheating temperature $T_R$ in
    the hybrid inflation model for the case 
    $M_1 \simeq 3 \times 10^9$ GeV ($a+d=2$).
    The contour lines are all shown by the dotted lines
    and corresponding values of $T_R$ are also represented
    in unit of GeV.
    The upper bound on $k$ from the requirement $\varphi_{N_e} < M_G$ is 
    shown by the thick solid line.
    The $k$ yielding $m_\varphi = m_\Sigma = 2M_1$ is also shown
    by the thick solid line.
    }
    \label{fig:TR_BL2}
\end{figure}

The reheating temperature $T_R$ depends on the mass of the lightest
Majorana neutrino, $M_1$ [see Eqs.  (\ref{eq:GAMW2-HInf}) and
(\ref{eq:TR-HInf})].  We show the $T_R$ in Figs. \ref{fig:TR_BL} and
\ref{fig:TR_BL2} for the case $M_1 \simeq 10^{12}$ GeV ($a+d=1$) and
for the case $M_1 \simeq 3 \times 10^9$ GeV ($a+d=2$), respectively.
%
%
\begin{figure}[t]
    \centerline{\psfig{figure=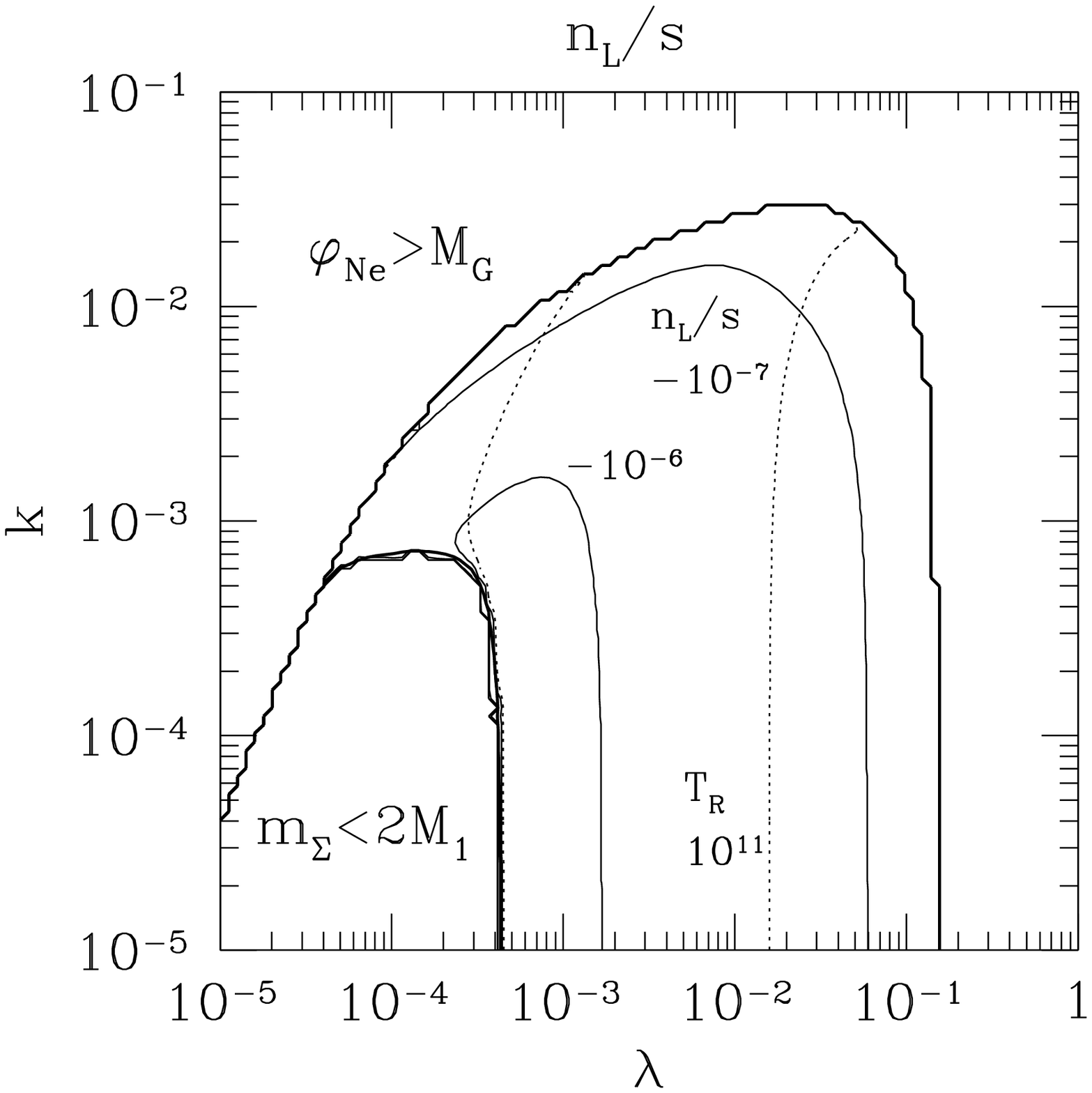,width=12cm}}
    \caption{
    The contour lines of the lepton asymmetry $n_L/s$ in
    the hybrid inflation model for the case 
    $M_1 \simeq 10^{12}$ GeV ($a+d=1$).
    The contour lines are all shown by the thin solid lines
    and corresponding values of $n_L/s$ are also represented.
    We also show the contour lines of the reheating temperature
    by the dotted lines and corresponding values of $T_R$ 
    are also represented in unit of GeV.
    The upper bound on $k$ from the requirement $\varphi_{N_e} < M_G$ 
    and the lower bound on $k$ from $m_\varphi = m_\Sigma 
    > 2 M_1$ are both shown by the thick solid lines.
    }
    \label{fig:LA_BL}
\end{figure}
%
\begin{figure}[t]
    \centerline{\psfig{figure=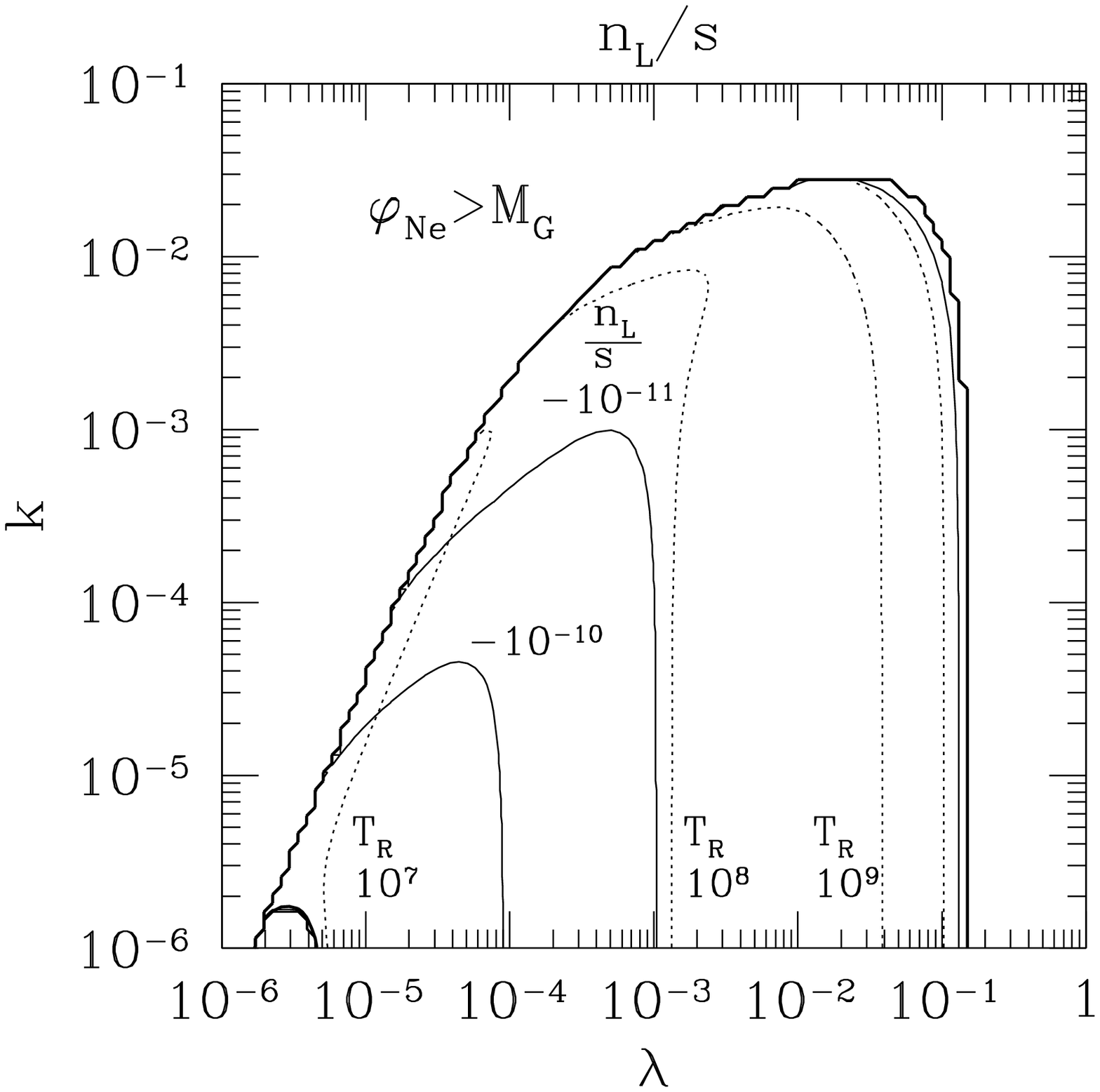,width=12cm}}
    \caption{
    The contour lines of the lepton asymmetry $n_L/s$ in
    the hybrid inflation model for the case 
    $M_1 \simeq 3\times 10^9$ GeV ($a+d=2$).
    The contour lines are all shown by the thin solid lines
    and corresponding values of $n_L/s$ are also represented.
    We also show the contour lines of the reheating temperature
    by the dotted lines and corresponding values of $T_R$ 
    are also represented in unit of GeV.
    The upper bound on $k$ from the requirement $\varphi_{N_e} < M_G$ 
    and the lower bound on $k$ from $m_\varphi = m_\Sigma 
    > 2 M_1$ are both shown by the thick solid lines.
    }
    \label{fig:LA_BL2}
\end{figure}
%
It is found that for the region of the inflaton mass $m_\varphi =
m_\Sigma \gg 2 M_1$ the reheating temperature $T_R \lesssim 10^8$ GeV,
which is required to avoid the cosmological gravitino problem, is
obtained only for the case $M_1 \simeq 3 \times 10^9$ GeV ($a+d=2$).
Notice that for the region $m_\varphi \le 2 M_1$ the reheating
temperature is determined by the decay width (\ref{eq:GAMK-HInf}) and
the desired low reheating temperature is obtained even for the case of
$M_1 \simeq 10^{12}$ GeV ($a+d=1$). However, such cases are not
interesting since the $N_1$ are not produced in the $\varphi$ and
$\Sigma$ decays and leptogenesis does not take place.

We now examine whether the leptogenesis works well or not in the above
hybrid inflation model.  Since the heavy Majorana neutrinos $N_1$ are
produced in the decays of the inflaton $\varphi$ and the $\Sigma$
field, the mass of the inflaton should satisfy Eq.
(\ref{eq:const-mphi}).  As derived in Sec.~\ref{sec:LA-InfDecay}, the
ratio of the produced lepton number to the entropy, $n_L/s$, is given
by Eq. (\ref{LA}).  Notice that $B_r \simeq 1$ in the present hybrid
inflation.  We show in Figs.  \ref{fig:LA_BL} and \ref{fig:LA_BL2} the
$n_L/s$ for the cases $a+d=1$ and $a+d=2$, respectively.
First, we consider the case of $M_1 \simeq 10^{12}$ GeV ($a+d=1$).  We
find from Fig. \ref{fig:LA_BL} that the lepton asymmetry enough to
explain the present baryon asymmetry can be generated in a wide
parameter region. However, we have too high reheating temperature of
$T_R \simeq 10^{9}$--$10^{12}$ GeV as mentioned before.%
\footnote{
For the higher reheating temperature of $T_R \gtrsim 10^{11}$ GeV,
the formula of the lepton asymmetry (\ref{LA}) should not be verified,
since $M_1 / T_R \lesssim 10$ and the inverse decay process
of the $N_1$ might be effective.
However, this does not affect our conclusion, because we 
concentrate ourselves only to the case $T_R \lesssim 10^8$ GeV.
}
(This parameter region may be interesting when some entropy
production of order $10^{3}$--$10^{4}$ takes place at low energies.)
Therefore, only the small 
region of $m_\Sigma \simeq 2 M_1$,
where the reheating temperature
becomes smaller as $T_R \simeq 10^{6}$--$10^8$ GeV due to the phase volume
suppression,%
\footnote{In this case, we should include the decay rates
(\ref{eq:GAMW1-HInf}) and 
(\ref{eq:GAMK-HInf}) in estimating $T_R$.  }
is free from the cosmological gravitino problem.  In this narrow region,
we obtain the required lepton asymmetry of 
$n_L/s \simeq - (0.3$--$3) \times10^{-10}$ 
[see Eq. (\ref{eq:LA-Req})]
to account for the present baryon asymmetry.

Next, we consider the case of $M_1 \simeq 3 \times 10^9$ GeV ($a+d=2$).
From Fig. \ref{fig:LA_BL2} it is found that 
the required lepton asymmetry of $n_L/s \simeq - 10^{-10}$ 
as well as the low enough reheating temperature of 
$T_R \simeq 10^{7}$--$10^{8}$ GeV%
\footnote{
In the region of $T_R \lesssim 10^{8}$ GeV
the formula for the lepton asymmetry Eq. (\ref{LA}) is justified
since $M_1 \gtrsim 10 T_R$.
}
are naturally offered in the region of $k \lesssim 10^{-3}$ and
$\lambda \simeq 10^{-6}$--$10^{-3}$.
Therefore, we conclude that the hybrid inflation with $M_1
\simeq 3 \times 10^9$ GeV can produce a sufficient lepton asymmetry
giving a reheating temperature low enough to solve the cosmological
gravitino problem \cite{Asaka-Hamaguchi-Kawasaki-Yanagida}.  However,
the reheating temperature of $T_R \simeq 10^6$ GeV which is required
to avoid the cosmological difficulty for the gravitino of mass
$m_{3/2} \simeq $100--500~GeV, is achieved only in the narrow parameter
region of $m_\varphi = m_\Sigma \simeq 2 M_1$ as in the previous case
($a+d=1$).

\clearpage

We have, so far, identified the $U(1)$ gauge symmetry in the hybrid
inflation model with the $B-L$ symmetry. 
We now consider the case where the $U(1)$ symmetry
is nothing related to the $B-L$ symmetry and even completely decoupled
from the SUSY standard-model sector.
The role of the $U(1)$ gauge symmetry is only to eliminate an
unwanted flat direction in Eq. (\ref{eq:W-HInf}).

\begin{figure}[t]
    \centerline{\psfig{figure=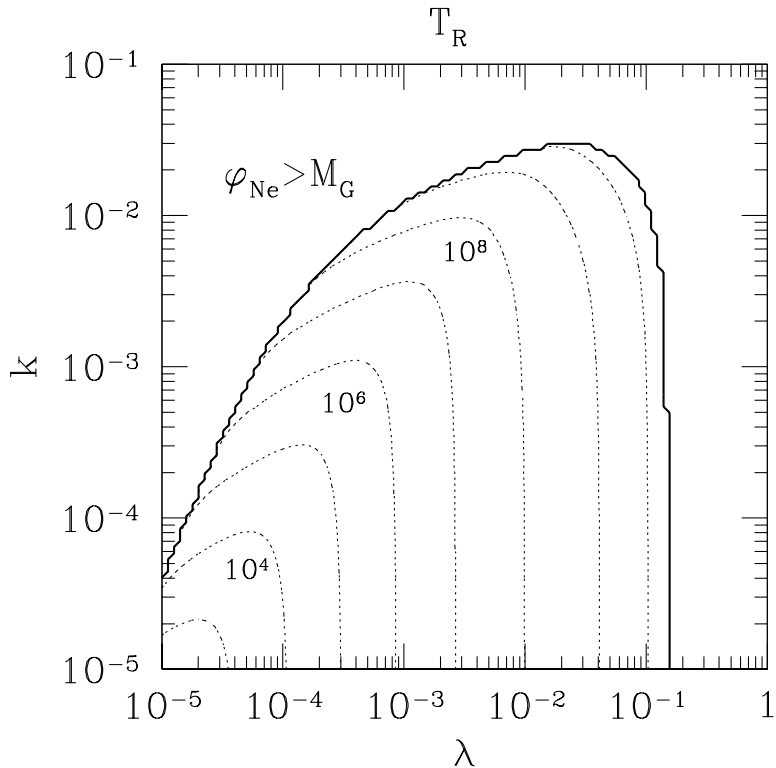,width=12cm}}
    \caption{
    The contour lines of the reheating temperature $T_R$ in
    the hybrid inflation model without the $B-L$ symmetry.
    The contour lines are all shown by the dotted lines
    and corresponding values of $T_R$ are also represented 
    in unit of GeV.
    The upper bound on $k$ from the requirement $\varphi_{N_e} < M_G$ is 
    shown by the thick solid line.
    }
    \label{fig:TR}
\end{figure}

\begin{figure}[ht]
    \centerline{\psfig{figure=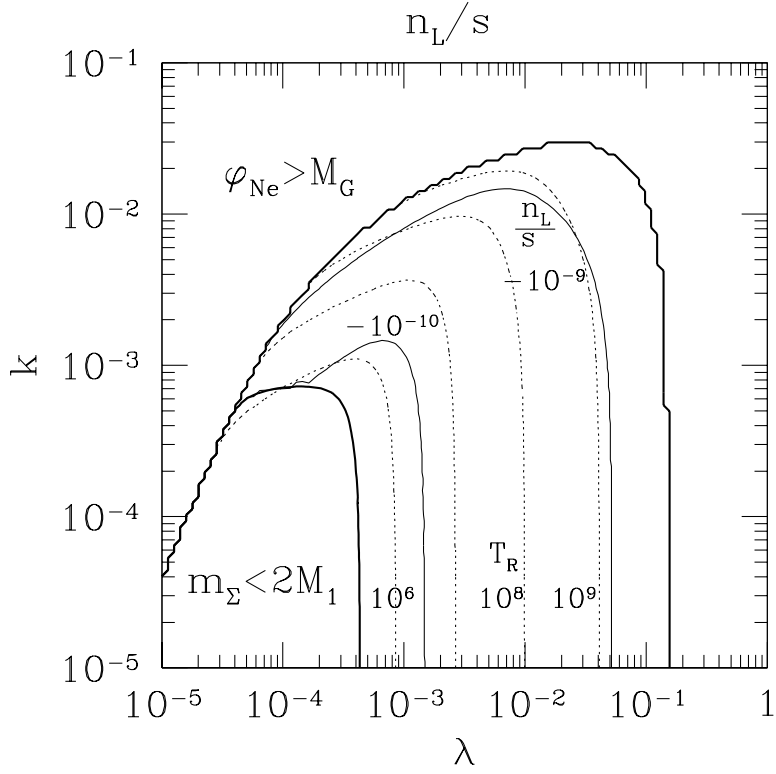,width=12cm}}
    \caption{
    The contour lines of the lepton asymmetry $n_L/s$ in
    the hybrid inflation model without the $B-L$ symmetry
    for the case $M_1 \simeq 10^{12}$ GeV ($a+d$=1).
    The contour lines are all shown by the thin solid lines
    and corresponding values of $n_L/s$ are also represented.
    Here we have assumed the branching ratio of 
    $\Sigma \rightarrow N_1 N_1$ decay channel $B_r = 1$
    for the estimation of $n_L/s$.
    We also show the contour lines of the reheating temperature
    by the dotted lines and corresponding values of $T_R$ 
    are also represented in units of GeV.
    The upper bound on $k$ from the requirement $\varphi_{N_e} < M_G$ 
    and the lower bound on $k$ from $m_\varphi = m_\Sigma 
    > 2 M_1$ are both shown by the thick solid lines.
    }
    \label{fig:LA}
\end{figure}

In this case the $\Sigma$ field may decay through the nonrenormalizable 
interactions (\ref{eq:KTR-HInf}).
On the other hand, the decay of the inflaton $\varphi$ is 
much suppressed due to the absence of the interaction
(\ref{eq:W-BL}). Thus, we introduce a new interaction
in the K\"ahler potential as
\footnote{
Without the K\"ahler potential (\ref{eq:KH-HInf}),
the inflaton $\varphi$ decays through interactions
in the K\"ahler potential $K \simeq |N_1|^2 |\Psi|^2 + |N_1|^2 
|\overline{\Psi}|^2$.
The rate of this inflaton decay into two $\widetilde{N_1}$
is estimated as
\begin{eqnarray}
    \label{eq:GAMK2-HInf}
    \Gamma (\varphi \rightarrow \widetilde{N_1} \widetilde{N_1} )
    \simeq \frac{1}{8 \pi} 
    \langle \Psi \rangle^2 M_1^2 m_\varphi.
\end{eqnarray}
In this case the inflaton decay is slower than 
the decay of the $\Sigma$ field since 
$m_\varphi=m_\Sigma > 2 M_1$
[see Eq. (\ref{eq:GAMK-HInf})], and the 
reheating temperature is determined by the decay rate
(\ref{eq:GAMK2-HInf}).
Then, the leptogenesis takes place in the inflaton decays
into two $\widetilde{N_1}$. (Note that $B_r \simeq 1$ is ensured.)
We find that the required lepton asymmetry is obtained
with lower reheating temperature such as $T_R \simeq 10^{6}$ GeV
for the region $\lambda \simeq 10^{-4}$--$10^{-3}$ and 
$k \lesssim 10^{-2}$ 
(see Fig. \ref{fig:LA}).
}
\begin{eqnarray}
    \label{eq:KH-HInf}
    K = h \phi^\ast H_u H_d + h.c.~.
\end{eqnarray}
(A natural symmetry allowing the K\"ahler potential 
(\ref{eq:KH-HInf}) will be discussed in the Appendix 
\ref{sec:appendix}.)
Through this interaction 
the inflaton $\varphi$ may decay faster than the $\Sigma$ field
for $h \simeq {\cal O}(1)$
\footnote{
We find $\Gamma_\varphi \gtrsim \Gamma_\Sigma$ 
for $h^2 \gtrsim C \langle \Psi \rangle^2
\simeq 10^{-6}$.
}
and the reheating temperature $T_R$ is given by 
the decay of the $\Sigma$ field.
Since the decay rate (\ref{eq:GAMK-HInf}) is very small
compared with Eq. (\ref{eq:GAMW2-HInf}), 
the reheating temperature $T_R$ 
becomes much lower than in the previous model.
The inflation dynamics is almost the same as in the previous hybrid
inflation. The difference of the results on various parameters, $\mu$,
$\langle \Psi \rangle$ and $m_\varphi$ from those in the previous
model may arise, in principle, from the different values of the reheating
temperatures $T_R$ as shown in Fig. \ref{fig:TR}.
However, we find no sizable difference in the obtained values of
$\mu$, $\langle \Psi \rangle$ and $m_\varphi$
in between the present and the previous models.

We turn to estimate the lepton asymmetry in this hybrid inflation model. 
We find that too small lepton asymmetry is
obtained for $M_1 \simeq 3 \times 10^9$ GeV ($a+d=2$) 
because of the small lepton-asymmetry parameter $\epsilon_1$
[see Eq. (\ref{eq:VEp1})],
and hence we concentrate our discussion to the case of 
$M_1 \simeq 10^{12}$ GeV ($a+d=1$) in the following analysis.
It is clear that the total decay rate (\ref{eq:GAMK-HInf}) of 
the $\Sigma$ field is much smaller than 
the decay rate of $N_1$ in Eq. (\ref{eq:Gam-N1}) and 
our approximation to derive Eq. (\ref{LA}) is justified.
%
We show the $n_L/s$ in Fig. \ref{fig:LA} assuming $B_r =1$.%
\footnote{
In this model, the $\Sigma$ field decays not only into the heavy Majorana
neutrinos $N_1$ but also into the SUSY standard-model particles
 through the nonrenormalizable interactions 
(\ref{eq:KTR-HInf}) and hence $B_r \simeq 1$ is not automatic.
}
It is found that the required lepton asymmetry to account for the
present baryon asymmetry can be generated in a wide parameter region
$k \lesssim 10^{-2}$ and $\lambda \simeq 10^{-3}$-- $10^{-2}$
with the reheating temperature of $T_R \simeq 10^{6}$--$10^{8}$ GeV.
It is extremely interesting that we obtain the desired result, $n_L/s
\simeq - 10^{-10}$ and $T_R \simeq 10^{6}$ GeV, for $\lambda \simeq
10^{-3}$ and $k \lesssim 10^{-3}$.  As shown in Ref.
\cite{Gravitino-Prob}, we may avoid the overproduction of gravitinos
in the full gravitino mass region of $m_{3/2} \simeq 100$ GeV-- 1 TeV
with such a low reheating temperature $T_R\simeq 10^{6}$ GeV.
Notice that the gravitinos are also produced in the reheating process
of the inflaton $\varphi$ decay.
However, they are diluted by the entropy production of the $\Sigma$ decays
and become negligible.
%
%
\clearpage
\section{Leptogenesis in New Inflation}
\label{sec:LG-NewInf}
As seen in the previous section, the SUSY hybrid inflation with $T_R
\simeq 10^{6}$--$10^8$ GeV is successful to produce the sufficient
lepton asymmetry to account for the baryon number in the present
universe.  However, we need small couplings, $k \lesssim 10^{-2}$ and
$\lambda \lesssim 10^{-2}$ to obtain the low reheating temperature
$T_R \simeq 10^6$--$10^8$ GeV.  Although this is not a problem in the
model, it is very important to find other inflation models
which naturally produce the required low reheating temperature.  The
new inflation \cite{New-Inflation} is a well-known candidate for such
inflation.  Therefore, we investigate, in this section, whether the
leptogenesis mechanism works in the new inflationary universe.

To perform a definite analysis we propose
a SUSY new inflation model,
\footnote{ Our discussion below can be also applied to the new
inflation model proposed by Izawa and Yanagida
\cite{Izawa-Yanagida-New}, which offers a scalar potential for
the inflaton field similar to Eq. (\ref{eq:V-NewInf}).
}
which has the following superpotential $W$ and K\"ahler potential $K$ for
supermultiplets
$\phi (x, \theta)$ and $\chi (x, \theta)$;
\begin{eqnarray}
  W 
  &=&
  \chi ( v^2 - g \phi^n )~, 
  \label{eq:W-NewInf}\\
  K 
  &=&
  |\phi|^2 + |\chi|^2 + \frac{\kappa_1}{4} |\phi|^4 + 
  \kappa_2 |\phi|^2|\chi|^2 + 
  \frac{\kappa_3}{4}|\chi|^4 \cdots ~,
  \label{eq:K-NewInf}
\end{eqnarray} 
where $g$, $\kappa_1$, $\kappa_2$ and  $\kappa_3$ are constants of order
unity, $v$ the energy scale of the inflation,
and the ellipsis denotes higher order terms.
The superpotential (\ref{eq:W-NewInf}) is naturally obtained,
for example, by imposing $U(1)_R\times Z_n$ symmetry.

From Eqs. (\ref{V-SUGRA}), (\ref{eq:W-NewInf}) and (\ref{eq:K-NewInf})
we find a SUSY vacuum,%
\footnote{
We always take $g$ and $v^2$ real and positive  by 
using the phase rotations of $\chi$ and $\phi$.
}
\begin{eqnarray}
    \label{eq:VEV-of-phi}
    \langle \phi \rangle = \left( \frac{v^2}{g} \right)^{ \frac{1}{n} }
    ~,
    \quad\langle \chi \rangle = 0 ~.
\end{eqnarray}
The potential of scalar fields for $\phi$, $\chi \ll 1$ is given by
\begin{eqnarray}
    V 
    \simeq v^4 
    - g v^2 \left( \phi^n + {\phi^{\ast}}^n \right)
    + g^2 |\phi|^{2n}
    + (1-\kappa_2) v^4 |\phi|^2 
    - \kappa_3 v^4 |\chi|^2,
\end{eqnarray}
where we have also used the same symbols $\phi$ and $\chi$ 
for the scalar components of
corresponding supermultiplets and neglected the higher order terms
represented by the ellipsis in Eq. (\ref{eq:K-NewInf}).

For $g>0$ and $k\equiv\kappa_2-1>0$, we can identify the inflaton
field $\varphi$ with the real part of the scalar field $\phi$.
Moreover, if $\kappa_3 < -\frac{3}{4}$, $\chi$ receives a mass which
is larger than $(3/2)H_I$ with the Hubble parameter $H_I$ during the
new inflation, and hence $\chi$ settles down at $\chi = 0$ quickly.
Hereafter, we assume $\kappa_3 < -\frac{3}{4}$ and set $\chi=0$.
Then, the inflaton potential near the origin ($\varphi \simeq 0$) is
obtained as
\begin{eqnarray}
 \label{eq:V-NewInf}
  V(\varphi) \simeq v^4 - \frac{k}{2}v^4\varphi^2
  - \frac{g}{ 2^{\frac{n}{2}-1} }v^2\varphi^n~.
\end{eqnarray}
The slow-roll conditions for the inflation Eqs. (\ref{eq:SR-Cond1})
and (\ref{eq:SR-Cond2}) are satisfied when
\begin{eqnarray}
    &&
    0 < k   \lesssim   1 ~,\\
    &&
    0 < \varphi   \lesssim  \varphi_f
    \equiv
    \sqrt{2}
    \left[
        \frac{ (1-k) v^2 }{ g n (n-1) }
    \right]^{ \frac{1}{n-2} }~.
\end{eqnarray}
Then, the new inflation takes place when the inflaton $\varphi$ rolls
down along the potential Eq. (\ref{eq:V-NewInf}) from $\varphi
\simeq 0$ to $\varphi_f$.
\footnote{
The initial value of the inflaton field $\varphi$ 
should be taken near the local maximum of the potential 
$\varphi \simeq 0$ to obtain a successful new inflation \cite{Linde}.
There seems no reason to take such an initial value,
since the potential should be flat enough for $\varphi$ 
to roll down slowly.
However, there have been found 
some dynamical mechanisms
\cite{Izawa-Kawasaki-Yanagida-NewInf,Asaka-Kawasaki-Yamaguchi}
to solve this initial value problem.
}

In the true vacuum Eq. (\ref{eq:VEV-of-phi}), the inflaton $\varphi$
has a mass as
\begin{eqnarray}
 \label{eq:mphi-NewInf}
  m_{\varphi} \simeq n g^{\frac{1}{n}} v^{2-\frac{2}{n}}~.
\end{eqnarray}
We assume that the inflaton decays through nonrenormalizable
interactions in the K\"ahler potential such as
\footnote{
Suitable charge assignment of $Z_n$ symmetry
for the Higgs supermultiplets $H_u$ and $H_d$ forbids 
a K\"ahler potential $K = \phi^* H_u H_d$.
}
\begin{eqnarray}
 \label{eq:KInfDecay-NewInf}
  K = \sum_i \lambda_i |\phi|^2|\psi_i|^2~,
\end{eqnarray}
where $\psi_i$ denote supermultiplets for SUSY standard-model particles
including the heavy Majorana neutrinos, and $\lambda_i$ are 
coupling constants of
order unity.
Then, the decay rate $\Gamma_{\varphi}$ is estimated as
\begin{eqnarray}
 \label{eq:Gam-NewInf}
 \Gamma_{\varphi}
  \simeq 
  \frac{1}{8\pi}
  \sum_i \lambda_i^2
  \langle \phi \rangle^2
  m_{\varphi}^3 ~,
\end{eqnarray}
which leads to
\begin{eqnarray}
 \label{eq:TR-NewInf}
  T_R 
  \simeq
  0.092 C' \langle \phi \rangle m_\varphi^{\frac{3}{2} } ~,
\end{eqnarray}
where $C'\equiv \sqrt{\sum{_i}|\lambda_i|^2}$ and we take $C'=1$
in the following discussion.
Then, the scale $v$ of the new inflation in Eq. (\ref{eq:W-NewInf})
is determined from the $e$-fold number $N_e$ of the present horizon,
the amplitude and the spectrum index $n_s$ of 
the primordial density fluctuations $\delta\rho/\rho$.

The number of $e$-foldings $N_e$ is given by Eq. (\ref{eq:Ne-PH}).
The present new inflation gives
\begin{eqnarray}
 N_e
  &=&
  \int^{\varphi_{N_e}}_{\varphi_f}
  d\varphi ~
  \frac{V(\varphi)}{V'(\varphi)} 
  \nonumber
  \\
 &\simeq&
  \int^{\varphi_{N_e}}_{\widetilde{\varphi}}
  d\varphi  ~
  \frac{v^4}{-kv^4\varphi}
  +
  \int^{\widetilde{\varphi}}_{\varphi_f}
  d\varphi ~
  \frac{v^4}{ -( ng/2^{ \frac{n}{2}-1 } ) v^2 \varphi^{n-1}  } ~,
  \label{eq:e-foldings}
\end{eqnarray}
where
\begin{eqnarray}
 \widetilde{\varphi}
 =
  \sqrt{2}
  \left(
   \frac{kv^2}{ng}
   \right)^{ \frac{1}{n-2} } ~,
\end{eqnarray}
and $\varphi_{N_e}$ is the value of the field $\varphi$ when the
present universe crossed the horizon.
Here we have assumed that $\varphi_{N_e} < \widetilde{\varphi}$ .%
\footnote{
This condition corresponds to $k \gtrsim 1/[N_e(n-2)]$.
}
Then, we find
\begin{eqnarray}
 \label{eq:varphi-N}
    \varphi_{N_e} \simeq
    \sqrt{2} \left( \frac{ k v^2 }{ ng } \right)^{\frac{1}{n-2}}
    \exp
    \left[
    - k 
    \left( N_e + 
        \frac{ n k - 1 }{ (n-2) k (1-k) }
    \right)
    \right]~.
\end{eqnarray}

The amplitude of the primordial density fluctuations
$\delta\rho/\rho$ due to the inflation should be normalized by the data
on anisotropies of the CMBR observed by
the COBE experiments 
as in the hybrid inflation.
From Eq. (\ref{eq:COBE-Norm}) we find
\begin{eqnarray}
 \label{eq:COBE-normalization}
 \frac{v^2}{ k\varphi_{N_e} }
  \simeq
  5.3 \times 10^{-4} ~.
\end{eqnarray}
Furthermore, the COBE observations show the spectrum index $n_s$ as 
$n_s = 1.0 \pm 0.2$
\cite{COBE}.
\footnote{
For the case of the scale invariant spectrum 
we have $n_s = 1$.
}
In the present new inflation model the index $n_s$ is calculated as 
\cite{Izawa-Yanagida-New}
\begin{eqnarray}
    \label{eq:n_s}
    n_s \simeq 1 - 2 k ~.
\end{eqnarray}
Therefore, we take $0.01 \lesssim k \lesssim 0.1$ in the following
analysis assuming the coupling $k$ not extremely small.
\footnote{
Notice that our assumption ($\varphi_{N_e} < \widetilde{\varphi}$)
in deriving Eq. (\ref{eq:e-foldings}) is
justified (see the footnote 21).  }

\begin{figure}[ht]
    \centerline{\psfig{figure=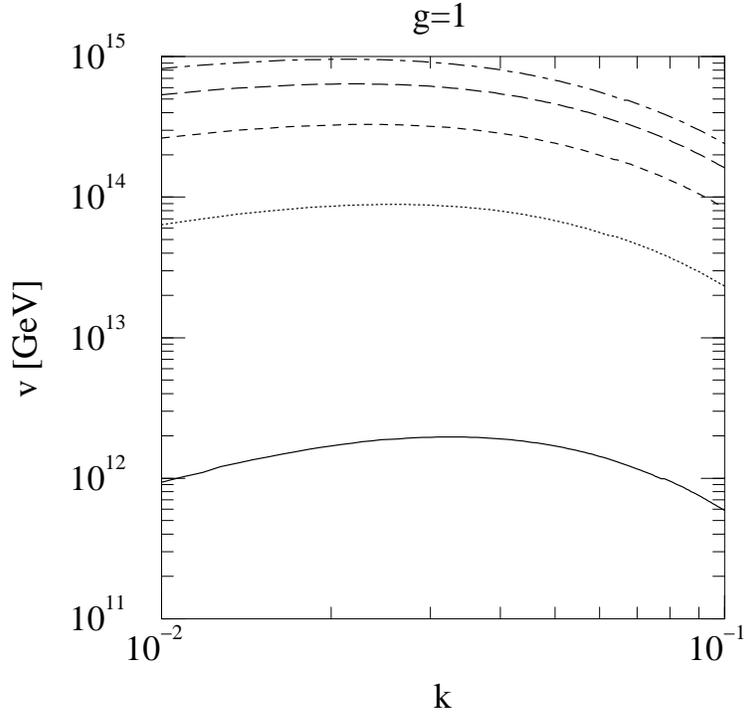,width=10cm}}
    \caption{
    The scale $v$ of the new inflation for $g=1$.
    We take the index $n$ as 
    $n=4$, 5, 6, 7 and 8 from the bottom to the top.
    }
    \label{fig:Vg=1}
\end{figure}

\begin{figure}[ht]
    \centerline{\psfig{figure=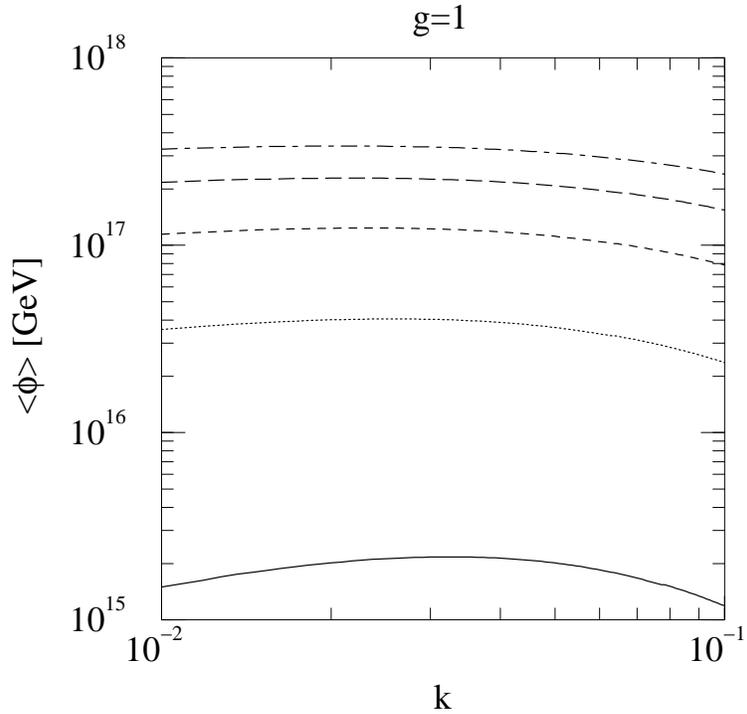,width=10cm}}
    \caption{
    The vev $\langle \phi \rangle$ of the new 
    inflation for $g=1$.
    We take the index $n$ as 
    $n=4$, 5, 6, 7 and 8 from the bottom to the top.
    }
    \label{fig:VEVg=1}
\end{figure}

\begin{figure}[ht]
    \centerline{\psfig{figure=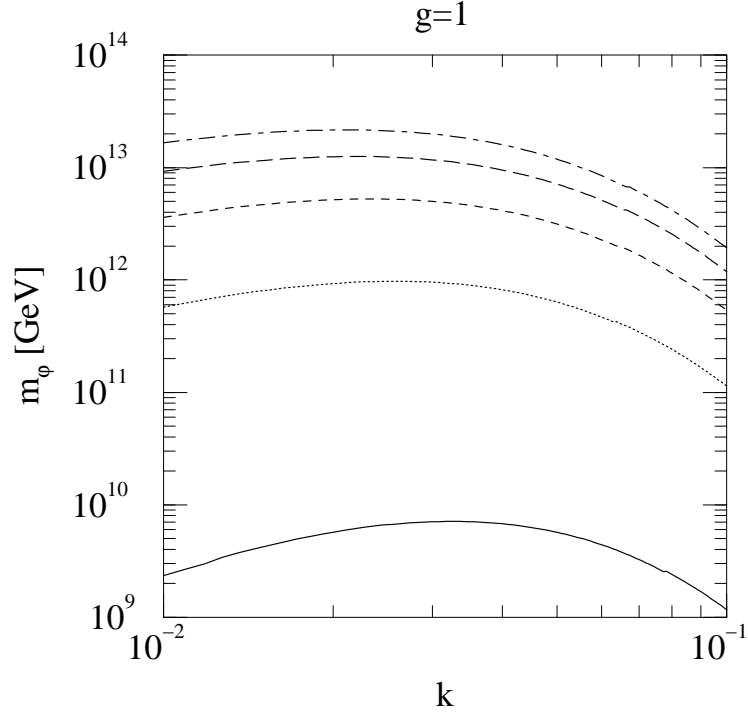,width=10cm}}
    \caption{
    The inflation mass $m_\phi$ of the new inflation for $g=1$.
    We take the index $n$ as 
    $n=4$, 5, 6, 7 and 8 from the bottom to the top.
    }
    \label{fig:MPHIg=1}
\end{figure}

\begin{figure}[ht]
    \centerline{\psfig{figure=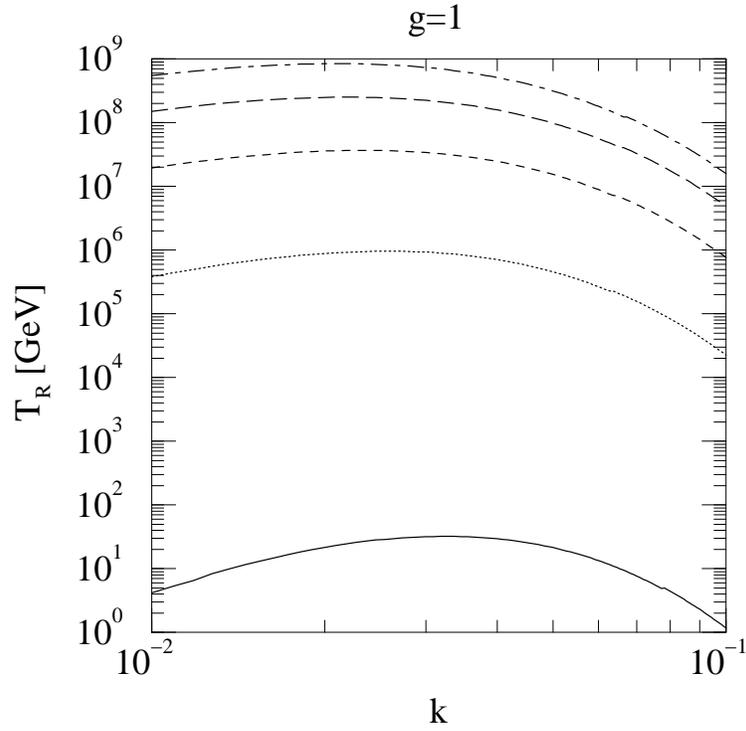,width=10cm}}
    \caption{
    The reheating temperature $T_R$ of the new inflation for $g=1$.
    We take the index $n$ as $n = $ 4, 5, 6, 7 and 8 from the bottom
    to the top.
    }
    \label{fig:TRg=1}
\end{figure}

We calculate the scale $v$ for given $n$, $g$ and $k$, which are
shown in Fig. \ref{fig:Vg=1}.  The vev $\langle
\phi \rangle$, the inflaton mass $m_{\varphi}$ and the reheating
temperature $T_R$ are found in Figs.  \ref{fig:VEVg=1},
\ref{fig:MPHIg=1} and \ref{fig:TRg=1}, respectively.  
From Fig.\ref{fig:TRg=1} 
we find that the new inflation model with the power index $n
=$ 4, 5 and 6 naturally offers the low reheating temperature
$T_R \lesssim 10^8$ GeV.
For the cases $n=$ 7 and 8, 
we obtain $T_R \lesssim 10^8$ GeV for the region
$k \simeq 0.05$--0.1 and 0.07--0.1, respectively.
\clearpage

We turn to discuss the leptogenesis in this new inflation model.
Here we also consider only the leptogenesis via
the decays of the lightest Majorana
neutrinos $N_1$ produced in the inflaton decay.
As well as in the previous hybrid inflation,
the inflaton decay rate $\Gamma_\varphi \simeq T_R^2$
is much smaller than the decay rate of $N_1$ in Eq. (\ref{eq:Gam-N1}),
and the decays of $N_1$ occur just after produced in the inflaton decay.
Thus, the ratio of the lepton-number density $n_L$ to the
entropy density $s$ is also given by Eq. (\ref{LA}).
Notice that the new inflation model with $n=4$ gives so low reheating
temperature as $T_R\simeq 1$ GeV--$10$ GeV (see
Fig. \ref{fig:TRg=1}), 
that the required amount of lepton asymmetry can not be
generated [see Eq. (\ref{LA})], and hence we discard this case.

\begin{figure}[t]
    \centerline{\psfig{figure=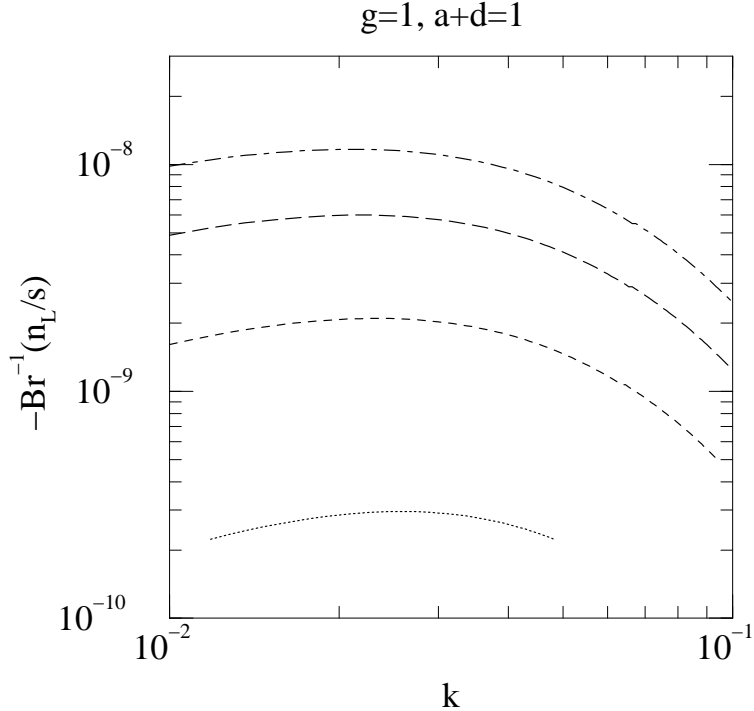,width=10cm}}
    \caption{
    The lepton asymmetry produced via the decays of $N_1$ in the new
    inflation.
    The index $n$ is taken as $n=5,6,7$ and $8$ from the bottom to the top.
    We take the FN charges $a$ and $d$ as $a+d=1$.
    For $n=5$, the regions $k < 1.2 \times 10^{-2}$ and
    $k > 4.8\times 10^{-2}$ are excluded since 
    $m_{\varphi}\le 2M_1$
    }
    \label{fig:LA_a+d=1}
\end{figure}
\begin{figure}[ht]
    \centerline{\psfig{figure=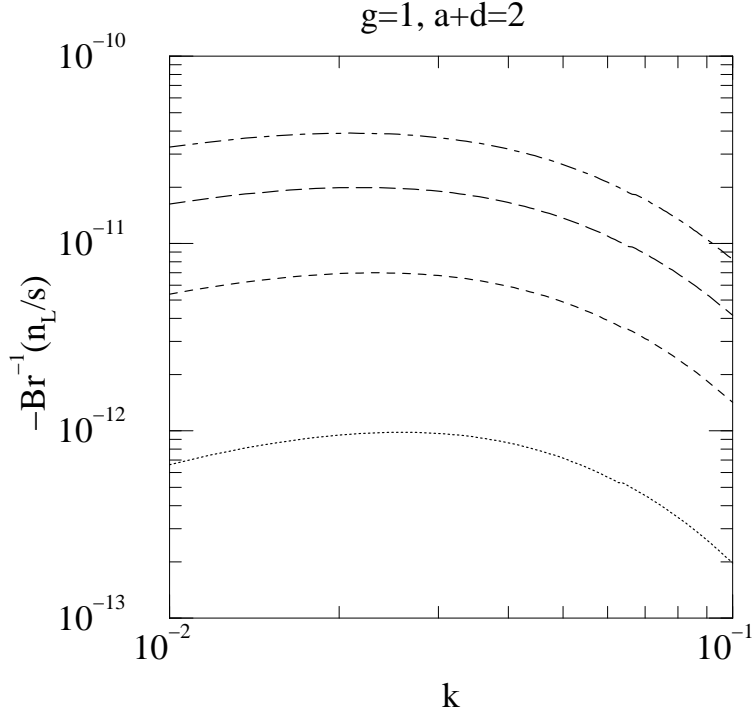,width=10cm}}
    \caption{
    The lepton asymmetry produced via the decays of $N_1$ in the new
    inflation.
    The index $n$ is taken as $n=5,6,7$ and $8$ from the bottom to the top.
    We take the FN charges $a$ and $d$ as $a+d=2$.
    }
    \label{fig:LA_a+d=2}
\end{figure}

For the case $n=5$, 6, 7 and 8, we show the produced lepton asymmetry in
Fig.~\ref{fig:LA_a+d=1} and Fig.~\ref{fig:LA_a+d=2} by taking $a+d=1$
and $a+d=2$, respectively.  
Here we have taken a smaller value of $M_1$ as
$M_1 \simeq \epsilon^{2(a+d)} 10^{14}$ GeV 
[see Eq. (\ref{eq:Mass-N1})] to obtain a wider allowed region,
taking account of ${\cal O}$(1) ambiguities in the FN model.

First, we consider the case $a+d=1$ and $M_1 \simeq 3 \times 10^{11}$
GeV.  In order to allow the decay $\varphi \rightarrow N_1 N_1$ the
inflaton mass should be larger than $2 M_1$, which excludes the
regions $k < 1.2 \times 10^{-2}$ and $k > 4.8\times 10^{-2}$ for the
case $n=5$.  We find from Fig. \ref{fig:LA_a+d=1} that the sufficient
lepton asymmetry can be generated for $n=5$, 6, 7 and 8 with the low enough
reheating temperature of $T_R\simeq 10^6$--$10^8$ GeV (see Fig.
\ref{fig:TRg=1}).  It should be noted that the required lepton
asymmetry $n_L/s\simeq - 10^{-10}$ is obtained for $T_R\simeq 10^6$
GeV when $B_r\simeq 1$ and $n=5$, and we can avoid the overproduction
of gravitinos even if $m_{3/2}\simeq 100$ GeV--500 GeV.  This is a
crucial result in the present analysis.  On the other hand, when $M_1
\simeq 10^9$ GeV (i.e., $a+d$=2), we find from Fig. \ref{fig:LA_a+d=2}
that the leptogenesis does not work well.%
\footnote{
For the case n=8, we obtain $B_r^{-1} (n_L/s) \simeq - 3 \times
10^{-11}$ in the region $10^{-2} \lesssim k \lesssim 0.4$. 
However, the reheating temperature is 
$T_R \gtrsim 10^8$ GeV and hence the cosmological gravitino problem
is still unsolved in this region.
}

%
\section{Leptogenesis in Topological Inflation}
\label{sec:LG-TopInf}
Finally, we discuss the leptogenesis in a topological inflation
\cite{Asaka-Hamaguchi-Kawasaki-Yanagida}. 
If the vev of the inflaton exceeds the gravitational scale, 
the inflaton potential (\ref{eq:V-NewInf}) in the new inflation model
is nothing but the inflaton potential for 
the topological inflation \cite{Top-Inf}.

For example, we consider the SUSY topological inflation model
proposed in Ref. \cite{Izawa-Kawasaki-Yanagida-TopInf}.
The superpotential and K\"ahler potential in the model
are given by
\begin{eqnarray}
    \label{eq:W-TopInf}
    W & = & v^2 \chi ( 1 - g \phi^2), \\
    \label{eq:K-TopInf}
    K & = & |\chi|^2 + |\phi|^2
    + \frac{ \kappa_1 }{ 4 } |\phi|^4
    + \kappa_2 |\chi|^2|\phi|^2 
    + \frac{ \kappa_3 }{4 }  |\chi|^4 + \cdots ~,
\end{eqnarray}
where $v$ is the energy scale of the inflation, $g$, $\kappa_1$,
$\kappa_2$ and $\kappa_3$ coupling constants of order unity.  These
potentials possess $U(1)_R \times Z_2$ symmetry, and the $U(1)_R$
charge of $\phi$ ($\chi$) is $-$2 (0) and $\phi$ ($\chi$) is odd
(even) under the $Z_2$.  This discrete symmetry is crucial for the
topological inflation.

From the scalar potential (\ref{V-SUGRA}),
and Eqs. (\ref{eq:W-TopInf}) and (\ref{eq:K-TopInf})
we find a SUSY-invariant vacuum,
\begin{eqnarray}
    \langle \chi \rangle = 0,
    \quad
    \langle \phi \rangle = \frac{1}{\sqrt{g}},
    \label{eq:VEVs-TopInf}
\end{eqnarray}
in which the potential energy vanishes.
The potential for the region $|\chi|$, $|\phi| \ll 1$
is written approximately as
\begin{equation}
    \label{eff-pot}
    V \simeq v^4|1 - g\phi^{2}|^{2} 
    + (1 - \kappa_2)v^4 |\phi|^2
    - \kappa_3 v^4 |\chi|^2 ~.
\end{equation}
The scalar components of
the supermultiplets are denoted by the same letters as the corresponding
supermultiplets.  
Hereafter, we set $\chi=0$ assuming $\kappa_3 < - 3/4$
as in the new inflation.
For $g > 0$ and $\kappa_2 < 1$, the inflaton field
$\varphi$ is identified with the real part of $\phi$ 
and the potential around the origin is given by
\begin{equation}
    \label{eff-pot2}
    V(\varphi) \simeq v^4 - \frac{k}{2} v^4 \varphi^2,
\end{equation}
where $k \equiv 2g + \kappa_{2} - 1$.
\footnote{
This is nothing but the potential (\ref{eq:V-NewInf})
in the new inflation model with $n=2$.
}
In the true vacuum (\ref{eq:VEVs-TopInf}) 
the inflaton mass is estimated as
\begin{equation}
    \label{eq:InfMass-TopInf}
    m_{\varphi} \simeq 2 \sqrt{g} v^2 
    = \frac{ 2 v^2 }{ \langle \phi \rangle } ~.
\end{equation}

A topological inflation takes place if the vev of $\phi$ is of 
order of the gravitational scale $M_G$ \cite{Top-Inf}.  As verified in
Ref. \cite{Sakai}, it should be
\begin{eqnarray}
    \label{eq:VEVcr-TopInf}
    \langle \phi \rangle 
    \ge \langle \phi \rangle_{cr}
    \simeq 1.7~.
\end{eqnarray}
The slow-roll conditions Eqs. (\ref{eq:SR-Cond1}) and
(\ref{eq:SR-Cond2}) are satisfied when $0 < k < 1$ and $\varphi 
\lesssim \varphi_f$ where $\varphi_f$ is expected to be of order of the
gravitational scale and we take $\varphi_f = 1$.
\begin{figure}[t]
    \centerline{\psfig{figure=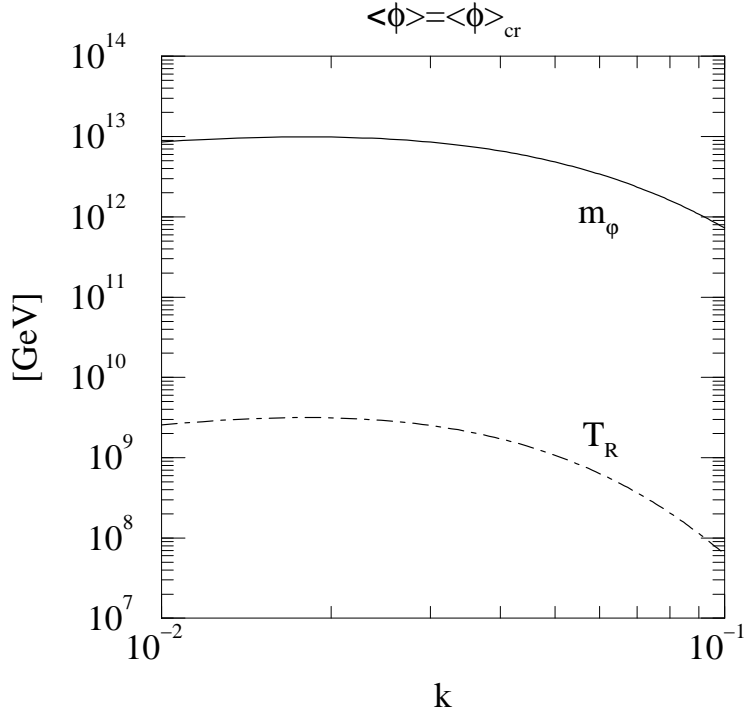,width=10cm}}
    \caption{
    The inflaton mass $m_\varphi$ and the reheating temperature $T_R$
    of the topological inflation model.
    We take $\langle \phi \rangle = \langle \phi \rangle_{cr} 
    \simeq 1.7 M_G$.
    The solid and dot-dashed lines represent $m_\varphi$ and $T_R$,
    respectively.
    }
    \label{fig:TopInf}
\end{figure}
\begin{figure}[ht]
    \centerline{\psfig{figure=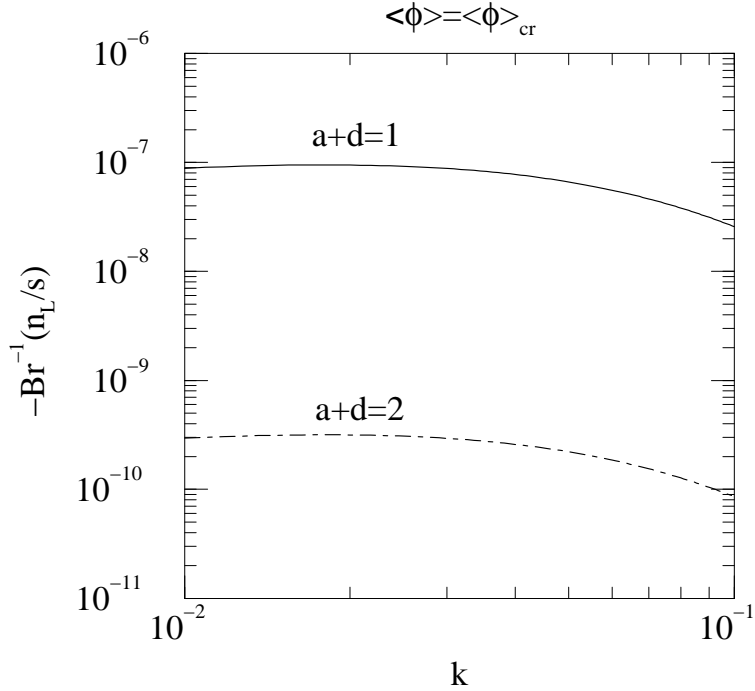,width=10cm}}
    \caption{
    The lepton asymmetries, $n_L/s$, in the topological inflation model
    for the case $a+d=1$ (the solid line) and 
    $a+d=2$ (the dot-dashed line).
    We take $\langle \phi \rangle = \langle \phi \rangle_{cr} 
    \simeq 1.7 M_G$.
    }
    \label{fig:LA_TopInf}
\end{figure}

Then, the scale of this topological inflation is determined by
the $e$-fold number $N_e$, the density fluctuations $\delta \rho/\rho$ 
and the spectrum index $n_s$.
The $e$-fold number can be written as
\begin{eqnarray}
    N_e 
    &\simeq&
    \int_{\varphi_f}^{\varphi_{N_e}} d\varphi ~
    \frac{V}{V'}
    \simeq
    \frac{ 1 }{ k } \ln 
    \left( \frac{ \varphi_f }{ \varphi_{N_e}} \right) ~.
\end{eqnarray}
We also assume that the inflaton $\varphi$ decays through 
the interactions in the K\"ahler potential as shown in 
Eq. (\ref{eq:KInfDecay-NewInf}) and hence the reheating
temperature of the topological inflation is also given by 
Eq. (\ref{eq:TR-NewInf}). 
$\delta \rho / \rho$ in the topological inflation can be 
calculated as
\begin{eqnarray}
    \frac{ \delta \rho }{ \rho }
    \simeq \frac{ 1 }{ 5 \sqrt{3} \pi } 
    \frac{ v^2 }{ k \varphi_{N_e} }~,
\end{eqnarray}
which should be normalized by the COBE data in Eq.
(\ref{eq:COBE-normalization}).  Furthermore, the spectrum index $n_s$
is given by $n_s \simeq 1 - 2 k$, and hence we take $0.01 \lesssim k
\lesssim 0.1$ as in the new inflation model.  In Fig. \ref{fig:TopInf}
we show the inflation mass $m_\varphi$ and the reheating temperature
$T_R$ of the topological inflation by taking $\langle \phi \rangle =
\langle \phi \rangle_{cr} \simeq 1.7$.  We see relatively high
reheating temperatures compared with those in the new inflation. This
is due to a larger value of $\langle \phi \rangle$.

Since the inflaton mass lies in the region $m_\varphi \simeq
10^{12}$--$10^{13}$ GeV and then the inflaton always decays into two
heavy Majorana neutrinos $N_1$ for the both $a+d=1$ and 2 cases.  Here
we also take the mass of the $N_1$ as $M_1 \simeq \epsilon^{2(a+d)}
10^{14}$ GeV [see Eq. (\ref{eq:Mass-N1})].  The inflaton decay rate
is much smaller than the decay rate of $N_1$ in Eq. (\ref{eq:Gam-N1})
and the estimation of the lepton asymmetry in Eq.~(\ref{LA}) is also
justified.  The ratio of the lepton number to the entropy density,
$n_L/s$, in the topological inflation is found in Fig.
\ref{fig:LA_TopInf}.  For the both $a+d=1$ and 2 cases, we obtain
lepton asymmetry $|n_L/s| \gtrsim 10^{-10}$
for $B_r \simeq 1$.  However, if
one requires the low reheating temperature in order to suppress
sufficiently the gravitino abundance as $T_R \lesssim 10^8$ GeV, the
coupling $k$ should be $k \gtrsim 0.092$.  (The lowest reheating
temperature $T_R = 6.3 \times 10^7$ GeV is obtained for $k=0.1$.)  For
$k\simeq 0.1$, the spectrum index $n_s$ deviates from the scale
invariant one as $n_s \simeq 0.8$, and it will be testable in future
satellite experiments~\cite{MAP} on anisotropies of the CMBR.

We should note finally that we may accommodate a late-time entropy
production of order $10^2$--$10^3$ in the case of $a+d=1$,
since the produced lepton asymmetry is large as 
$n_L/s \simeq - 10^{-7}$ as shown in Fig. \ref{fig:LA_TopInf}.
If it is the case, the energy density of the gravitinos is diluted 
by the factor $10^2$--$10^3$, which allows the region of
$T_R \simeq 10^9$ GeV in Fig. \ref{fig:TopInf}.
\section{Discussion and Conclusions}
\label{sec:discussion}
%
We have discussed, in this paper, the leptogenesis via the decays of
the heavy Majorana neutrinos $N_1$ produced non-thermally in the
inflaton decay. We have performed a detailed analysis on the
leptogenesis taking three types of SUSY models for hybrid, new and
topological inflations, and found that all of them are successful to
produce the lepton-number asymmetry enough to account for the baryon
asymmetry in the present universe. Here we have imposed the reheating
temperatures $T_R \lesssim 10^{8}$ GeV to suppress sufficiently the
density of gravitinos for keeping the success of the BBN for $m_{3/2}
\simeq {\cal O}(1)$ TeV.  However, much lower reheating temperature
such as $T_R \simeq 10^{6}$ GeV is required \cite{Gravitino-Prob} for
$m_{3/2} \simeq 100$ GeV--500 GeV, which is only achieved in the
hybrid or new inflation model.  In other words, the leptogenesis is
fully consistent with the BBN in a wide range of the gravitino mass
of $m_{3/2} \simeq 100$
GeV--1 TeV in the hybrid or new inflationary universe.
It should be noted that
the leptogenesis in these inflation models are also 
consistent with a class of the gauge-mediated SUSY breaking models
\cite{GMSB},
because the overclosure problem of the stable gravitino
of $m_{3/2} \simeq 10$ MeV--1 GeV can be avoided 
when the reheating temperature is $T_R \simeq 10^6$ GeV
\cite{Moroi-Murayama-Yamaguchi}.

The above hybrid inflation with $T_R \simeq 10^6$ GeV is realized by
taking small couplings $k \lesssim 10^{-3}$ and $\lambda \simeq
10^{-3}$ (see Fig.~\ref{fig:LA}), which predicts the breaking scale of
the $U(1)$ gauge symmetry at $\langle \Psi \rangle = \langle
\overline{\Psi} \rangle \simeq (4$--$6)\times 10^{15}$ GeV. This
$U(1)$ breaking produces cosmic strings in the early universe.  The
cosmic strings with such a high breaking scale 
of $10^{15}$--$10^{16}$
GeV are very interesting, since they give additional contributions to
the anisotropies of the CMBR~\cite{String}, which may be testable in
future satellite experiments~\cite{MAP}.
We should note that the scale-invariant spectrum
index ($n_s \simeq 1$) is also predicted in the hybrid inflation model.

The successful new inflation with $T_R \simeq 10^{6}$ GeV is given
by taking the $k \simeq 0.01$--$0.05$ (see Figs. \ref{fig:TRg=1}
and \ref{fig:LA_a+d=1}), which yields the spectrum index
$n_s \simeq 0.98$--$0.90$. We expect that some parameter region
of $n_s$ may be excluded (or confirmed) in the future experiments
\cite{MAP}.

In the present paper, we have assumed the FN model for mass matrices
of quarks and leptons. However, it is straightforward task to apply
the present analysis to another model for Yukawa couplings for quarks
and leptons to evaluate the asymmetry parameter $\epsilon_1$.
The leptogenesis with more generic Yukawa matrices will be 
discussed in a future publication~\cite{Asaka-Hamaguchi-Kawasaki-Yanagida2}.
%
\section*{Acknowledgements}
%
The authors would like to thank to Toshiyuki Kanazawa and Masahide
Yamaguchi for useful discussion.  This work was partially supported by
the Japan Society for the Promotion of Science (TA,KH) and ``Priority
Area: Supersymmetry and Unified Theory of Elementary Particles
($\sharp$707)''(MK,TY).
\clearpage
\appendix
\section{}
\label{sec:appendix}
%
In this appendix, we show a symmetry which naturally provides 
the superpotential (\ref{eq:W-HInf}) as well as the K\"ahler potential
(\ref{eq:KH-HInf}) for the inflation
supermultiplet $\phi (x, \theta)$.
It is very crucial to suppress $W = \lambda' \phi H_u H_d$ 
in the hybrid inflation model.

We adopt $U(1)_R$ $R$--symmetry and $R$--charges of relevant
supermultiplets are found in Table \ref{tab:Rcharges}.
\begin{table}[t]
    \begin{center}
    \begin{tabular}{| c | c c c  | c c  | c c c | }
              & $\phi$ & $\Psi$ & $\overline{\Psi}$ &
        $H_u$ & $H_d$  & $l_i$  & $e^c_i $ & $N_i$\\
        \hline
        $Q_R$    & 2     & 0    &  0     &
        $2-2a$  & $3-3a$ &  $2a-1$  & $a$ & 1 \\
    \end{tabular}
    \caption{The $U(1)_R$ charges of various supermultiplets.}
    \label{tab:Rcharges}
    \end{center}
\end{table}
%
We see that the dangerous superpotential $W = \lambda' \phi H_u H_d$
is forbidden by choosing $a \neq 1$.

Now we discuss, by taking $a=0$ for example, a breaking of $U(1)_R$
to induce a K\"ahler potential (\ref{eq:KH-HInf}), 
which is very important to cause the 
inflaton $\varphi$ decay in the second hybrid inflation model
as explained in the text. We assume that the $U(1)_R$ symmetry
is broken down to $Z_3$-symmetry by the vev of the $\Xi$ supermultiplet
carrying a $U(1)_R$-charge $+3$. Because of the positive $R$-charge
of $\Xi$ the superpotential does not receive any $U(1)_R$-breaking
effects. However, we may have the following K\"ahler potential:
\begin{eqnarray}
    K = f \Xi^\ast \phi^\ast H_u H_d + h.c.~.
\end{eqnarray}
Then, the non-vanishing vev of the $\Xi$ field gives rise to
Eq. (\ref{eq:KH-HInf}) and 
the coupling constant $h$ in Eq. (\ref{eq:KH-HInf}) is given
by
\begin{eqnarray}
    h = f \langle \Xi^\ast \rangle~.
\end{eqnarray}
%

\end{document}